\documentclass[aps,prd,superscriptaddress,nofootinbib,11pt]{revtex4}
\usepackage[english]{babel}
\usepackage[utf8]{inputenc}
\usepackage{graphicx}   
\usepackage{slashed}
\usepackage{epstopdf}
\usepackage{verbatim}   
\usepackage{color}      
\usepackage{subfigure}  
\usepackage{multirow}
\usepackage{hyperref}   
\usepackage{float}
\usepackage{epsfig,rotating}
\usepackage{amsmath,amssymb}
\usepackage{dsfont}
\usepackage{slashed}
\restylefloat{table}
\numberwithin{equation}{section}

\newcommand{\vx}{\vec{x}}

\newcommand{\vk}{\vec{k}}

\newcommand{\be}{\begin{equation}}
\newcommand{\ee}{\end{equation}}
\newcommand{\bea}{\begin{eqnarray}}
\newcommand{\eea}{\end{eqnarray}}

\newcommand{\ket}[1]{|#1\rangle}
\newcommand{\bra}[1]{\langle#1|}

\newcommand{\walpha}{\widetilde{\alpha}}
\newcommand{\wbeta}{\widetilde{\beta}}
\newcommand{\wgamma}{\widetilde{\gamma}}

\newcommand{\wR}{\widetilde{\mathbb{R}}}
\newcommand{\wD}{\widetilde{D}}
\newcommand{\wP}{\widetilde{P}}

\begin{document}
\title{Synthetic and cosmological axion hybridization: \\ entangled photons, (HBT) and quantum beats.}

\author{Daniel Boyanovsky}
\email{boyan@pitt.edu} \affiliation{Department of Physics, University of Pittsburgh, Pittsburgh, PA 15260}

 \date{\today}

\begin{abstract}

In this article it is argued that synthetic axions, emergent collective excitations in topological insulators or Weyl semimetals  hybridize  with the cosmological axion, a compelling dark matter candidate via a common two photon decay channel since they both couple to electromagnetic fields via a Chern-Simons term. We point out an analogy to a V-type three level system with the two upper levels identified with the synthetic and cosmological axions decaying into a two-photon state. The Weisskopf-Wigner theory of spontaneous decay in multi level atoms is complemented and extended  to describe the dynamics of hybridization. The final two-photon state features both kinematic and polarization entanglement and displays quantum beats as a consequence of the interference between the decay paths.  An initial population of either axion induces a population of the other via hybridization. Consequently, a dark matter axion condensate induces a condensate of the synthetic axion, albeit with small amplitude.   We obtain a momentum and polarization resolved Hanbury- Brown Twiss (HBT) second order coherence describing coincident correlated two-photon detection. It exhibits   quantum beats with a frequency given by  the difference between the energies of the synthetic and cosmological axion and   \emph{perhaps may be harnessed} to detect either type of axion excitations.    The case of   synthetic axions individually is obtained in the limit of vanishing coupling of the cosmological axion and features similar two-photon correlations. Hence   second order (HBT) two-photon coherence  \emph{may} provide an alternative detection mechanism for emergent condensed matter axionic collective excitations. Similarities and differences with parametrically down converted photons are discussed.

\end{abstract}


\maketitle

\section{Introduction}\label{sec:intro}

The axion is a hypothetical elementary particle that was originally  introduced in Quantum Chromodynamics (QCD) as a solution of the strong CP problem\cite{PQ,weinaxion,wil} and is   a potentially viable  cold dark matter candidate\cite{pres,abbott,dine}. Extensions beyond the standard model of particle physics may include pseudoscalar particles with properties similar to the QCD axion and  can also be suitable dark matter candidates\cite{banks,ringwald,marsh,sikivie1}. While the cosmological axion has not yet been detected, various experiments are   focused on its detection  \cite{cast,admx,graham,irastorza}.   An important feature that characterizes the axion field $\phi(\vx,t)$  is  its pseudoscalar nature and its interaction with  photons  via pseudoscalar composite operators of gauge fields, such as $\vec{E}\cdot\vec{B}$ in the form of a Chern-Simons term in the Lagrangian density;
\be \mathcal{L}_{CS}= g_a \phi(\vx,t) \vec{E}(\vx,t)\cdot \vec{B}(\vx,t) \,, \label{Cster}\ee  defining the burgeoning field of axion electrodynamics\cite{wilczekaxion}.

Remarkably, in condensed matter physics  axion quasiparticles emerge as collective excitations in magnetic topological insulators\cite{xiao,nomura,rundong,jing,narang} or as  axionic charge density waves in Weyl semimetals\cite{gooth,yu,mottola} or in multilayered metamaterials\cite{wilczekshapo}. These collective excitations  couple to electromagnetism via a Chern-Simons term, just as the cosmological axion, leading to  topological magnetoelectric effects such as Faraday and Kerr rotations\cite{liang,tse,ahn}, with potential impact in   axion detection\cite{ishi}. Whereas there are no experimental confirmations of axionic collective excitations in condensed matter systems yet, recently it has been proposed that the vanishing of the bulk gap at the surface of magnetic topological insulators yield an enhancement of the fluctuations of the axion field near the surface providing a possible platform for their detection\cite{zaletel}. Recent studies\cite{marsh7,jan,so}, proposed to exploit the Primakoff effect, axion-photon conversion in presence of an external magnetic field, to probe a condensate of the cosmological dark matter axion with   synthetic axion quasiparticle excitations in topological magnetic insulators.

\vspace{1mm}

\textbf{Main  objectives:}

Our study is motivated by the observation that the couplings of the synthetic ($\phi_s$) and cosmological ($\phi_c$) axions to electromagnetism via the  Chern-Simons term (\ref{Cster}) entails that both types emit and absorb two photons, and as a consequence of this common   channel they hybridize via the process $\phi_s \Leftrightarrow 2 \gamma \Leftrightarrow \phi_c$. Secondly, we realize that there is a compelling analogy to a V-type three level system\cite{zubairy,meystre} with the upper two levels  decaying into a lower state by spontaneous emission.
In this case, if an initial state is prepared in a coherent superposition of the two upper levels, the final state exhibits quantum beats as a consequence of the path interference in the emission process\cite{zubairy,meystre}. The analogy with   synthetic/cosmological axions  emerges when the    upper two levels are identified with these excitations   and the decay is via a two-photon process. The nature of the Chern-Simons coupling suggests that the two-photon final state is entangled both in momentum and \emph{polarization}. Consequently, the analogy to a V-type system leads us to   conclude that the amplitude of the final correlated two-photon state,  will   exhibit quantum beats   with potentially interesting observational consequences. The correlations in the final two photon state can be revealed by   the Hanbury-Brown Twiss (HBT) second order coherence including polarizations, which should exhibit quantum beat phenomena and \emph{may} provide an experimentally accessible platform to confirm  the hybridization between the synthetic and cosmological axions via coincident  photon detection.

We emphasize that the hybridization and dynamical phenomena that our study focuses on are entirely different from the synthetic-cosmic axion coupling via axion-photon conversion in presence of external magnetic fields (Primakoff effect) invoked in references\cite{marsh7,jan,so}.  The hybridization phenomena that we   study is independent of any  external magnetic field.  It  is a direct consequence of a common two-photon decay channel of synthetic and cosmic axion (quasi) particle excitations, which leads to  an off-diagonal one-loop self-energy matrix determined by the two-photon intermediate state.

Our objectives with this study are the following: \textbf{i:)} to study the \emph{dynamics} of synthetic-cosmic axion hybridization by exploiting the analogy to a V-type three level system. For this purpose we extend and complement the formulation of references\cite{meystre,garraway,plenio} of spontaneous emission in multilevel atoms to the quantum field theory of synthetic and cosmological axions. \textbf{ii:)} Using this framework we obtain the  amplitude of the two photon final state  resulting from the decay of the respective axion fields, focusing on extracting quantum beat phenomena that results from the interference of the decay channels. \textbf{iii:)} Decay kinematics from the Chern-Simons interaction term determines that this two-photon final state will be entangled both in momentum and in \emph{polarization} as a consequence of the pseudoscalar nature of the coupling. We   introduce   a momentum and polarization resolved  Hanbury-Brown Twiss (HBT) second order (intensity) coherence  to study two-photon correlations including polarization. \textbf{iv:)} finally, we recognize a striking analogy to parametric down conversion, suggesting that intensity interferometry (HBT) is also a useful probe of correlations in the case of axions, in particular providing a possible alternative probe of axions in condensed matter platforms.

\vspace{1mm}

\textbf{Brief summary of results:}

\textbf{i:)} We recognize that the coupling of synthetic and cosmological axions to electromagnetism via a Chern-Simons term implies a common two-photo decay channel which implies hybridization between the two species via the mutual absorption and emission of photons as intermediate states: $\mathrm{synthetic ~ axion} \Leftrightarrow 2 \gamma \Leftrightarrow \mathrm{cosmological~ axion}$. This process implies an \emph{off diagonal self energy}.

\textbf{ii:)} A similarity with a ``V''-type three-level atom is established, with the two upper levels identified with the synthetic and cosmological axions respectively, decaying to a ground level via two-photon spontaneous emission.  The Weisskopf-Wigner theory of   spontaneous emission in multilevel atoms is adapted to study the dynamics of hybridization and to obtain the amplitude of the two-photon state. The kinematics of decay and the pseudoscalar nature of the Chern-Simons couplings entail that the photons are entangled both in momentum and polarization with a distinct polarization pattern. We obtain the amplitude of this two photon state to leading order in the couplings. As a direct consequence of hybridization, it is found that if the initial amplitude of one of the axion fields vanishes, the other axion field generates a non-vanishing amplitude during the time evolution. As a corollary, a condensate of the cosmological axion as a dark matter component, induces a condensate of the synthetic axion, albeit with small amplitude determined by the couplings.

\textbf{iii:)} To probe both momentum and polarization entanglement as a telltale signal of axion hybridization, we introduce a momentum and polarization resolved Hanbury-Brown Twiss (HBT) second order coherence (or cross correlation) as a \emph{possible} experimental tool to study the telltale photon correlations via coincident two-photon detection.  It features quantum beats as a consequence of the interference between the decay paths with a beat frequency determined by the
difference of the (renormalized) synthetic and cosmological axion single particle energies.

\textbf{iv:)} We argue that the same (HBT) correlations may be useful to probe condensed matter axions only, even when the small coupling of the cosmic axion to electromagnetism entails unobservable quantum beat phenomena. By taking the limit of vanishing cosmological axion coupling, it follows that the axion collective excitations in topological insulators and or Weyl semimetals may also be probed via the momentum and polarization resolved (HBT) second order coherence because the photons from the synthetic axion decay feature the same type of correlations, determined by their coupling to electromagnetism. Therefore, we propose that (HBT) second order correlations \emph{maybe} a useful tool to probe synthetic axions.

\textbf{v:)} We recognize a striking similarity to the case of parametrically down converted photons discussing similarities and differences. This similarity suggests that experimental interferometric methods to study parametric down conversion \emph{may} also be suitable to study emergent axionic excitations in condensed matter system via their two-photon decay channel with distinct momentum and polarization correlations.

In section (\ref{sec:axions}) we introduce the effective field theory for synthetic and cosmological axions, and discuss the interaction picture to study time evolution to leading order in the couplings. Section (\ref{sec:mixing}) presents the extension of Weisskopf-Wigner theory that describes the dynamics of spontaneous emission in multilevel atoms to study the non-equilibrium dynamics of hybridization and to obtain the amplitude of the two photon final state. In section (\ref{subsec:hybrid}) the general results are applied to the case of synthetic and cosmological axion hybridization, establishing momentum and polarization entanglement and that interference between the decay paths leads to quantum beats in the probability amplitude of the two photon state. In this section  we introduce the momentum and polarization resolved (HBT) second-order coherence (or cross correlation) as a useful experimental probe to study hybridization  via coincident photo-detection. Momentum and polarization entanglement with a particular polarization pattern is a telltale of axion electrodynamics and (HBT) correlations \emph{maybe} important useful probes both in the case of hybridization or solely for condensed matter axion excitations as discussed in detail in section (\ref{subsec:cmaxion}). In this section  we also discuss the striking similarity to parametric down conversion suggesting that (HBT) interferometry could also be a useful probe to study axion physics in condensed matter systems.

In section (\ref{sec:discussion}) we discuss various aspects and caveats of the approximations involved in the analysis.  Our conclusions and further avenues of study are summarized in section (\ref{sec:conclusions}). Various appendices are devoted to technical aspects.

\section{Synthetic and cosmic axions.}\label{sec:axions}
The effective action for the emergent (synthetic) dynamical axion field in topological insulators is given by\cite{rundong,jing,nomura}\footnote{We have absorbed the vacuum dielectric constant, vacuum permittivity  and a factor $4\pi$ into a redefinition of the gauge fields and set the speed of light $c=1$. }
\be  {S}_s = \int d^3x dt \Bigg\{ \frac{1}{2}\Big(  \vec{E}^2- {\vec{B}^2}  \Big)+\frac{\mathcal{J}}{2} \Bigg(\Big(\frac{\partial \delta \theta}{\partial t}\Big)^2- \Big(\vec{v}\cdot \vec{\nabla}\delta \theta\Big)^2-m^2_s\, \delta \theta^{\,^2} \Bigg) + \frac{\alpha}{ \pi} \,\delta \theta \,\vec{E}\cdot \vec{B}  \Bigg\}\,,\label{synac}  \ee where  the synthetic dynamical axion field $\delta \theta$ is related to the fluctuations of the Neel order parameter in the case of topological magnetic insulators, $\mathcal{J},\vec{v}$ are model and material dependent constants, $m_s$ is the synthetic axion mass\cite{rundong,jing,nomura}, and  $\alpha$ is the fine structure constant. Redefining the (canonically normalized) synthetic axion field
\be \phi_s(\vx,t) = \sqrt{\mathcal{J}}\, \delta \theta(\vx,t) \,,\label{physint}\ee
the effective action for the synthetic axion field and the electromagnetic field is
\be {S}_s = \int d^3x dt \Bigg\{ \frac{1}{2}\Big(  \vec{E}^2- {\vec{B}^2}  \Big)+\frac{1}{2} \Bigg(\Big(\frac{\partial \phi_s(\vx,t)}{\partial t}\Big)^2- \Big(\vec{v}\cdot \vec{\nabla}\phi_s(\vx,t)\Big)^2-m^2_s\, \phi^{\,2}_s(\vx,t)  \Bigg) + g_s \,\phi_s(\vx,t) \,\vec{E}\cdot \vec{B}  \Bigg\} \,,\label{synacfin} \ee with
$  g_s \equiv  {\alpha}/{\pi\sqrt{\mathcal{J}}}$.  While the form of the second bracket in $S_s$ may differ in the realizations of synthetic dynamical axions in topological insulators and Weyl semimetals, the coupling to electromagnetism via the last, Chern-Simons,  term is a generic and distinct hallmark of the coupling of synthetic axions to the $U(1)$ gauge fields.

\begin{figure}[ht]
\includegraphics[height=2.5in,width=2.5in,keepaspectratio=true]{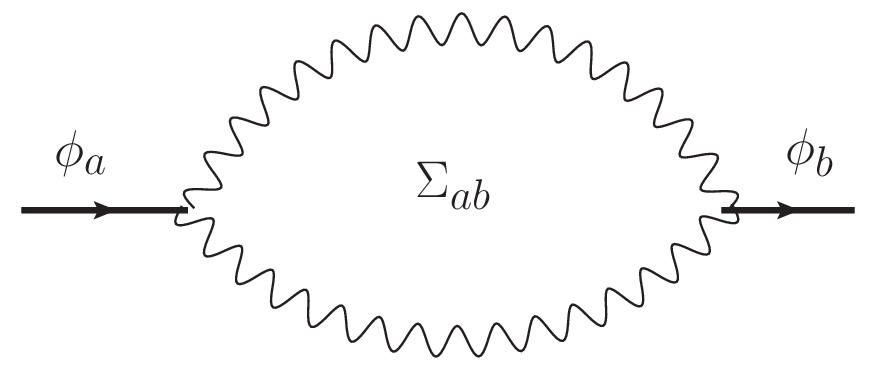}
\caption{ Synthetic (s) and cosmological (c) axion hybridization via the off-diagonal one loop self-energy diagram with two-photon exchange arising from the correlation function of the Chern-Simons density $\vec{E}\cdot \vec{B}$. The labels $a,b = s,c$. }
\label{fig:selfenergy}
\end{figure}

The effective action for the cosmological axion field $\phi_c(\vx,t)$ interacting with electromagnetism is  given by\cite{PQ,weinaxion,wil,wilczekaxion}
\be {S}_c = \int d^3x dt \Bigg\{ \frac{1}{2}\Big(  \vec{E}^2- {\vec{B}^2}  \Big)+\frac{1}{2} \Bigg(\Big(\frac{\partial \phi_c(\vx,t)}{\partial t}\Big)^2- \Big( \vec{\nabla}\phi_c(\vx,t)\Big)^2-m^2_c\, \phi^{\,2}_c(\vx,t)  \Bigg) + g_c \,\phi_c(\vx,t) \,\vec{E}\cdot \vec{B}  \Bigg\} \,.\label{cosmoac} \ee Therefore we consider the total action that describes together the synthetic and cosmological axions
\be S_{tot} = S_{0EM}+S_{0s}+S_{0c}+ \int d^3x dt \,\Big(g_s\phi_s(\vx,t)+g_c \phi_c(\vx,t)\Big)\,\vec{E}(\vx,t)\cdot\vec{B}(\vx,t) \,,\label{Stotal}\ee with $S_{0EM},S_{0s},S_{0c}$ are  the free-field actions for electromagnetism,  synthetic and cosmological axions respectively. Writing the total action in this form makes manifest that the interactions of the synthetic and cosmological axions with the electromagnetic fields lead to emission/absorption  processes $\phi_a \Leftrightarrow 2 \gamma \Leftrightarrow \phi_b$ with a two-photon intermediate state and $a,b \equiv s,c$.
These processes define the  one-loop self-energy matrix $\Sigma_{ab}, a,b = s,c$ depicted in fig. (\ref{fig:selfenergy}) from which the off diagonal elements $\Sigma_{sc},\Sigma_{cs}$ describe  the mixing or hybridization of the synthetic and cosmological axions.

One of our main objectives is to derive the self-energy matrix and to study the dynamics  of hybridization.

 The time evolution   is determined by the quantum Hamiltonian operator. Since the interaction involves derivative terms, Hamiltonian quantization is
subtle  because the Chern-Simons interaction modifies the canonical momentum of the gauge field. The proper quantization procedure along with the interaction picture representation are  discussed in detail in appendix (\ref{app:hamcs}).    The main result is that  to leading order in the interaction the total Hamiltonian is
\be H = H_{0EM}+H_{0s}+H_{0c}+H_I \,,\label{totHam}\ee with $H_{0EM},H_{0s},H_{0c}$ are the free field Hamiltonians for electromagnetism, synthetic and cosmological axions respectively and to leading order in the couplings $g_s,g_c$     the interaction Hamiltonian  in  \emph{interaction picture} is given by
\be H_I(t) = - \int d^3x \Big(g_s\phi(\vx,t)+g_c\phi_c(\vx,t)\Big)\vec{E}(\vx,t)\cdot\vec{B}(\vx,t) \,,\label{HIoft2}\ee where (see appendix   (\ref{app:hamcs}))
\be \vec{E}(\vx,t) = - \frac{\partial}{\partial t}\vec{A}(\vx,t) \,. \label{efield}\ee

The quantized fields in   interaction picture are expanded as
 \be \phi_{s,c}(\vx,t) = \frac{1}{\sqrt{V}}\,\sum_{\vk} \frac{1}{\sqrt{2E_{s,c}(k)}}\,\Big[b_{s,c}(\vk)\,e^{-iE_{s,c}(k)t}\,e^{i\vk\cdot \vx} + b^\dagger_{s,c}(\vk)\,e^{iE_{s,c}(k)t}\,e^{-i\vk\cdot \vx}  \Big]\,,\label{scalarquant2}\ee where $V$ is the quantization volume and
 \be E_s(k) = \sqrt{\Big(\vec{v}\cdot\vec{k}\Big)^2+m^2_s}~~;~~E_c = \sqrt{k^2+m^2_c} \,,\label{energies} \ee are the single (quasi) particle energies for the synthetic (s) and cosmological (c) axions respectively. In Coulomb gauge the vector potential is given by
  \be \vec{A}(\vx,t) =  \frac{1}{\sqrt{V}}\,\sum_{\vk}\sum_{\lambda=1,2}  \frac{\vec{\epsilon}_\lambda(\vk)}{\sqrt{2k}}\, \Big[a_{\lambda}(\vk)\,e^{-ik t}\,e^{i\vk\cdot \vx} + a^\dagger_{\lambda}(\vk)\,e^{ikt}\,e^{-i\vk\cdot \vx}  \Big]\,,\label{vecpot2}\ee where the annihilation and creation operators obey the usual canonical commutation relations and the real linear polarization vectors $\vec{\epsilon}_{\lambda}(\vec{k})$ are defined in appendix (\ref{app:hamcs}) (see equation (\ref{polas})).

\section{ Weisskopf-Wigner theory of hybridization. :}\label{sec:mixing}

We begin by  extending the  formulation of spontaneous emission of multi-level atoms\cite{meystre,garraway,plenio} based on the Weisskopf-Wigner theory of atomic linewidths\cite{ww,zubairy,meystre},   to the case when excitations of different species hybridize via a common set of intermediate states, or common decay channel. The method parallels the effective field theory of flavor mixing phenomena developed in ref.\cite{shuyang}. We   first discuss the general case, and apply it to the hybridization of synthetic and cosmic axions in the next section.

Consider a system whose Hamiltonian $H$ is given as a soluble part $H_0$ and a perturbation $H_I$: $H=H_0+H_I$. The time evolution of states in the interaction picture
of $H_0$ is given by
\be i \frac{d}{dt}|\Psi(t)\rangle_I  = H_I(t)\,|\Psi(t)\rangle_I,  \label{intpic}\ee
where the interaction Hamiltonian in the interaction picture is
\be H_I(t) = e^{iH_0\,t} H_I e^{-iH_0\,t} \label{HIoft}\,,\ee where $H_I$ is proportional to a set of couplings assumed to be small.

 The state $|\Psi(t)\rangle_I$ can be expanded as  \be |\Psi(t)\rangle_I = \sum_n C_n(t) |n\rangle \label{decom}\ee where $|n\rangle$ form a complete set of orthonormal states chosen to be eigenfunctions of $H_0$, namely $H_0\ket{n} = E_n\ket{n}$; in the many body case these are  many-particle Fock states. From eqn.(\ref{intpic}), and the expansion (\ref{decom})  one finds the   equation of motion for the coefficients $C_n(t)$, namely

\be \dot{C}_n(t) = -i \sum_m C_m(t) \langle n|H_I(t)|m\rangle \,. \label{eofm}\ee

Although this equation is exact, it generates an infinite hierarchy of simultaneous equations when the Hilbert space of states spanned by $\{|n\rangle\}$ is infinite dimensional. However, this hierarchy can be truncated by considering the transition between states connected by the interaction Hamiltonian at a given order in $H_I$.

\begin{figure}[ht]
\includegraphics[height=3.0in,width=3.0in,keepaspectratio=true]{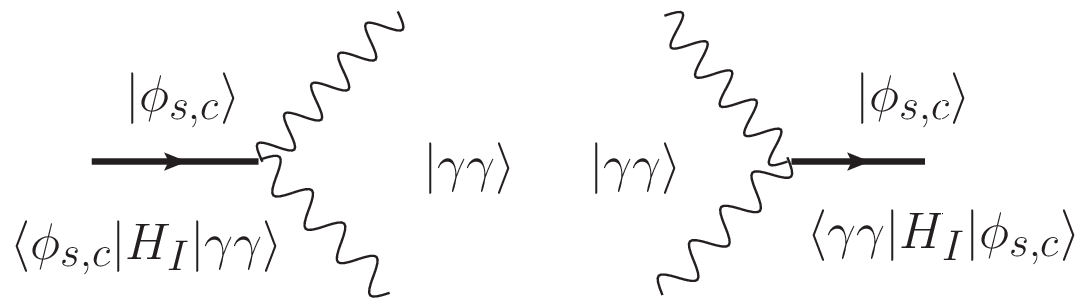}
\caption{Hybridization between $\ket{\phi_s},\ket{\phi_c}$ mediated by a common set of two-photons intermediate states $\ket{\kappa}=\ket{\gamma\gamma}$ with the interaction Hamiltonian  $H_I = -\int  (g_s\phi_s+g_c\phi_c)\,\vec{E}\cdot\vec{B}\,d^3x$. }
\label{fig:transitions}
\end{figure}

Let us consider quantum states $\ket{\phi_s},\ket{\phi_c}$ associated with the synthetic ($s$) and cosmological $(c)$ axion  fields  respectively, these may be single particle momentum eigenstates of the Fock quanta of these fields, and focus on  the case when the interaction Hamiltonian does not couple \emph{directly} the states $\ket{\phi_s},\ket{\phi_c}$, namely $\langle\phi_{s,c}|H_I|\phi_{s,c}\rangle =0$. Instead these states are connected to a common set of intermediate states $|\{\kappa\}\rangle$ by  $H_I$, namely $\ket{\phi_{s,c}} \Rightarrow |\{\kappa\}\rangle \neq \ket{\phi_{s,c}}$. In the case under consideration the  states $\ket{\phi_{s,c}}$ are single (quasi) particle excitations of the synthetic and cosmological axion fields, and the intermediate state $|\kappa\rangle$ is a two-photon state created by the operator $\vec{E}\cdot \vec{B}$ as depicted in fig. (\ref{fig:transitions}).

 The states $\ket{\phi_s},\ket{\phi_c}$ hybridize as a consequence of this indirect coupling through the common set of intermediate states, namely $\ket{\phi_{s,c}} \Leftrightarrow |\{\kappa\}\rangle \Leftrightarrow \ket{\phi_{s,c}}$, yielding an off-diagonal self-energy matrix.

In the   subspace $\ket{\phi_s},\ket{\phi_c},  |\{\kappa\}\rangle$ the quantum state in the interaction picture is given by
\be |\Psi\rangle_I(t) = \Big(C_s(t)\ket{\phi_s}+ C_c(t) \ket{\phi_c}\Big)\otimes \ket{\{0\}_\kappa}+ \sum_{\{\kappa\}} C_{\kappa}(t)  \ket{\kappa}\otimes\ket{0_s,0_c}   \,, \label{state}\ee
and the set of equations (\ref{eofm})   become\footnote{The interaction Hamiltonian also yields contributions in which the initial particle is unaffected and two photons and a synthetic or cosmological axion are produced. These are non-energy conserving ``vacuum type'' diagrams that will be neglected.}
\bea
\dot{C}_{s}(t) & = & -i \sum_{\{\kappa\}}  \langle\phi_s|H_I(t)| \kappa \rangle \, C_{\kappa}(t)\, \label{eqnc1} \\
\dot{C}_{c}(t) & = & -i \sum_{\{\kappa\}}  \langle\phi_c|H_I(t)| \kappa \rangle \, C_{\kappa}(t)\,, \label{eqnc2} \\
\dot{C}_{ \kappa}(t) & = & -i  \Bigg[\langle  \kappa |H_I(t)|\phi_s\rangle\, C_{ s}(t)+ \langle  \kappa |H_I(t)|\phi_c\rangle\,  C_{ c}(t)\Bigg]\,.\label{eqncm}\eea   where   the time dependent transition matrix elements are given by
\be \langle l|H_I(t)|m\rangle = T_{lm}\,  e^{i(E_l-E_m)t} ~~;~~ T_{lm}= \langle l|H_I(0)|m\rangle \,,\label{mtxele} \ee hermiticity of $H_I$ entails that
  \be T_{ml} = T^*_{lm} \,.\label{hermi}\ee

  The set of equations (\ref{eqnc1}-\ref{eqncm}), truncates the hierarchy of equations by neglecting the transitions between the states $|\{\kappa\}\rangle$ and  $|\{\kappa'\}\rangle \neq |\{\kappa\}\rangle,\ket{\phi_{1,2}}$, such transitions connect the states $\ket{\phi_{s,c}} \leftrightarrow |\{\kappa'\}\rangle $ at higher order in $H_I$ and are neglected up to $\mathcal{O}(H^2_I)$.  Truncating the hierarchy closes the set of equations for the amplitudes, effectively reducing the set of states to a closed subset in the full Hilbert space.

  Taking the normalized initial  quantum state $ \ket{\Psi(t=0)}$ as a coherent linear superposition of the single particle states $\ket{\phi_{s,c}}$, it is given by
\be \ket{\Psi(t=0)} = \Big(C_{s}(0)\ket{\phi_s} + C_{c}(0)\ket{\phi_c}\Big)\otimes \ket{0_\kappa}  \,,\label{inistate} \ee where $\ket{0_\kappa}$ is the vacuum state for the intermediate states $\ket{\kappa}$, in the case under consideration, it corresponds to the electromagnetic vacuum, corresponding to setting
\be C_{\kappa}(0) = 0 \,, \label{cmini}\ee  and with normalization condition
\be   |C_{s}(0)|^2+ |C_{c}(0)|^2  =1 \,. \label{unit}  \ee

\begin{figure}[ht]
\includegraphics[height=3.0in,width=3.0in,keepaspectratio=true]{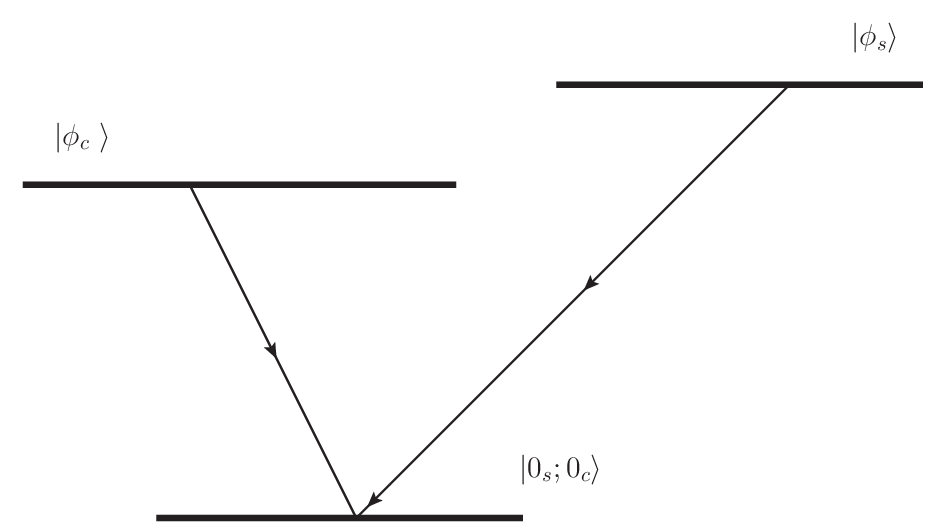}
\caption{Equivalence to a ``V-type'' atomic system with two upper levels decaying to a common lower level by spontaneous emission. }
\label{fig:vtype}
\end{figure}

This system is strikingly similar to a ``V-type'' system of three atomic levels with the two upper levels decaying to a common lower level by spontaneous emission, depicted in fig.(\ref{fig:vtype}) where the two upper levels are identified with the synthetic and cosmological axion excitations.

This type of systems is ubiquitous in quantum optics\cite{zubairy,meystre,garraway,plenio} and gives rise to quantum beats as a consequence of  ``which path'' interference.

\vspace{1mm}

\textbf{Unitarity:} The set of equations (\ref{eqnc1}-\ref{eqncm}) describe \emph{unitary time evolution} in the restricted Hilbert space of states $\ket{\phi_s},\ket{\phi_c},|\kappa\rangle$ which is a sub-set of the full Hilbert space of the theory that is closed under the equations  of motion (\ref{eqnc1}-\ref{eqncm}). Unitarity can be seen as follows: using the equations (\ref{eqnc1}-\ref{eqncm}), and noticing that $\langle l|H_I(t)|m\rangle^* = \langle m|H_I(t)|l\rangle $ because $H_I(t)$ is an Hermitian operator, it follows from equations (\ref{eqnc1}-\ref{eqncm}) that
\be \frac{d}{dt} \Bigg[ | {C}_{s}(t)|^2+   | {C}_{c}(t)|^2 + \sum_{\{\kappa\}} | {C}_{\kappa}(t)|^2\Bigg] =0 \,,\label{totder}\ee and the initial conditions (\ref{unit},\ref{cmini}) yield \be |{C}_{s}(t)|^2+   | {C}_{c}(t)|^2 + \sum_{\{\kappa\}} | {C}_{\kappa}(t)|^2 = 1 \,.  \label{unitarity} \ee This is the statement that time evolution within the sub-Hilbert space $\Big\{\ket{\phi_s},\ket{\phi_c},|\kappa\rangle\Big\}$  is unitary.

In particular if the $\phi_{s,c}$ states decay, it follows that $|C_{s,c}(t=\infty)|^2=0$,  and asymptotically
\be \sum_{\kappa}|C_{\kappa}(t=\infty)|^2 = 1 \,. \label{inftyti}\ee

The set of equations (\ref{eqncm}) with the initial condition (\ref{cmini}) can be integrated to yield
\be C_\kappa(t) = -i \int^t_0 \Bigg[T_{\kappa s} \,e^{i(E_\kappa-E_s)t'}\, C_{ s}(t')+ T_{\kappa c} \,e^{i(E_\kappa-E_c)t'}\,  C_{ c}(t')\Bigg]\,dt' \,,\label{cmoft} \ee where the labels $s,c$ correspond to $\phi_{s,c}$. Inserting the solution  (\ref{cmoft}) into the equations (\ref{eqnc1},\ref{eqnc2}) leads to the following set of equations for the amplitudes $C_{s}(t),C_{c}(t)$
\be
  \dot{C}_{s}(t)   =  -  \int^t_0  \sum_{\kappa}\Bigg\{   |T_{s\kappa}|^2 \, e^{i(E_s-E_\kappa)(t-t')}\,   {C}_{s}(t') +  T_{s\kappa}T_{\kappa c}\,e^{i(E_1-E_c)t}\,e^{i(E_c-E_\kappa)(t-t')} {C}_{c}(t') \Bigg\} \, dt'\,, \label{dotC1} \ee
\be \dot{C}_{c}(t)   =   -  \int^t_0  \sum_{\kappa}\Bigg\{  T_{c\kappa}T_{\kappa s}\,e^{i(E_c-E_s) t}\,e^{i(E_s-E_\kappa)(t-t')} {C}_{s}(t') + |T_{c\kappa}|^2\,e^{i(E_c-E_\kappa)(t-t')}\,    {C}_{c}(t') \Bigg\}  \, dt' \,. \label{dotC2}
\ee

This procedure of solving for the amplitudes of  the intermediate states   plays the role of  ``tracing over'' the $\kappa$ degrees of freedom,   yielding an effective set of equations of motion for the amplitudes of the single particle states $\ket{\phi_{s,c}}$. Since the interaction Hamiltonian $H_I$ is assumed to be proportional to weak couplings, the amplitude equations (\ref{dotC1},\ref{dotC2}) are exact up to second order in these couplings.

\subsection{Markov approximation: the effective non-Hermitian Hamiltonian}\label{subsec:markov}

Let us define
\be \int^{t'}_0 \sum_{\kappa} T_{a\kappa}T_{\kappa b}\,  e^{i(E_b-E_{\kappa})(t-t'')}\,dt''    \equiv W_{ab}[t;t']~~;~~ W_{ab}[t;0]=0 \,\label{defW}\ee so that

\be \sum_{\kappa} T_{a\kappa}T_{\kappa b} e^{i(E_b-E_{\kappa})(t-t')}= \frac{d}{dt'}W_{ab}[t;t']\,,\label{iden} \ee where the   labels $a,b$ stand for $s$ (synthetic) or $c$ (cosmic). Inserting this definition in (\ref{dotC1},\ref{dotC2}) and integrating by parts
\be \int^t_0 \frac{d}{dt'}W_{ab}[t;t']\,C_b(t')\,dt' = W_{ab}[t;t]\,C_b(t)-\int^t_0  W_{ab}[t;t']\,\frac{d}{dt'}C_b(t')\,dt'\,,\label{intparts}\ee since $T_{a\kappa} T_{\kappa b} \propto H^2_I$ and from the evolution equations (\ref{dotC1},\ref{dotC2}) it follows that $\dot{C}_a \propto H^2_I$ therefore,  the second term on the right hand side in (\ref{intparts}) is of $\mathcal{O}(H^4_I)$ and will be neglected to leading order in the interaction, namely $H^2_I$.

 Hence, up to $\mathcal{O}(H^2_I)$, the evolution equations for the amplitudes (\ref{dotC1},\ref{dotC2}) become
\bea \dot{C}_s(t) & = & - \Big\{W_{ss}[t;t]\,C_s(t)+ e^{i(E_s-E_c)\,t}\,W_{sc}[t;t]\,C_c(t)    \Big\}\,\label{dotc1fin} \\
\dot{C}_c(t) & = & - \Big\{e^{i(E_c-E_s)\,t}\,W_{cs}[t;t]\,C_s(t)+  W_{cc}[t;t]\,C_c(t)    \Big\}\,.\label{dotc2fin}
\eea Introducing  the Schroedinger picture amplitudes
\be e^{-iE_s t}\,C_s(t)\equiv A_s(t) ~~;~~ e^{-iE_c t}\,C_c(t)\equiv A_c(t)\,,\label{As}\ee

  the amplitude equations for $A_{s,c}$ become
\bea i\,\dot{{A}}_s(t) & = &  E_s A_s(t) - i\,W_{ss}[t;t]A_s(t)-i\,W_{sc}[t;t]A_c(t) \,\label{amp1} \\
 i\,\dot{{A}}_c(t) & = &  E_c A_c(t) - i\,W_{cs}[t;t] A_s(t)-i\,W_{cc}[t;t]A_c(t) \,.\label{amp2}\eea

\begin{figure}[ht]
\includegraphics[height=3.0in,width=3.0in,keepaspectratio=true]{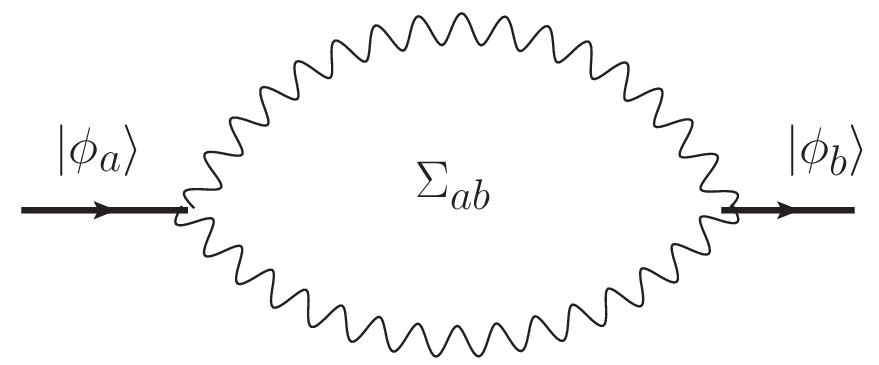}
\caption{One loop self-energy diagram corresponding to the two-photon exchange arising from the correlation function of the Chern-Simons density $\vec{E}\cdot \vec{B}$ that hybridizes the synthetic (s) and cosmic (c) axions. The labels $a,b = s,c$. }
\label{fig:onelup}
\end{figure}

Following the main Markov approximation in the Weisskopf-Wigner method\cite{ww,zubairy} we  invoke the
 long time limit
 \be \int^{t}_0    e^{i(E_b-E_{\kappa})(t-t' )}\,dt'~{}_{\overrightarrow{t\rightarrow \infty }}~ i\,\Bigg[\mathcal{P}\,\Big(\frac{1}{E_b-E_\kappa} \Big) -i \pi \delta(E_b-E_\kappa) \Bigg]\,,\label{pp} \ee and let us introduce     the spectral densities
\be \rho_{ab}(k_0) =  \,\sum_{\kappa} T_{a \kappa} T_{\kappa b} \,\delta(k_0-E_{\kappa}) = \rho^*_{ba}(k_0)~~;~~ a,b = s,c \,, \label{rhosd}\ee where the second identity follows from equation (\ref{hermi}),  yielding
 \be -i\,W_{ab}[t;t]~{}_{\overrightarrow{t\rightarrow \infty }}~\Sigma_{ab}(E_b) \,. \label{Ws} \ee The self-energy
 \be \Sigma_{ab}(E_b)  = \int^{\infty}_{-\infty} \frac{\rho_{ab}(k_0)}{E_b-k_0+ i\varepsilon} \,dk_0   ~~;~~ \varepsilon \rightarrow 0^+ \label{selfenergy}\ee has a simple and intuitive interpretation as a second order Feynman diagram wherein the lines representing the  intermediate states $\ket{\kappa}$ in fig. (\ref{fig:transitions}) are joined into ``propagators'' yielding a  one loop diagram,   representing the self-energy up to second order in $H_I$. For the case under consideration this Feynman diagram features a two-photon loop arising from the correlation function of the Chern-Simons density $\vec{E}\cdot \vec{B}$ and is  depicted in figure (\ref{fig:onelup}).

This is precisely what was anticipated in the introduction, confirming the intuitive discussion on hybridization via the common decay channel in section (\ref{sec:axions}).

Separating the self-energy into the real and imaginary parts
\be  \Sigma_{ab}(E_b)  = \Delta_{ab}(E_b)-i \frac{\Gamma_{ab}(E_b)}{2} ~~;~~ \Delta_{ab}(E_b) \equiv \int^{\infty}_{-\infty}\mathcal{P} \Bigg[  \frac{\rho_{ab}(k_0)}{E_b-k_0} \Bigg]\,dk_0 ~~;~~ \Gamma_{ab}(E_b) = 2\pi\,\rho_{ab}(E_b)\,,\label{RIsig}\ee where $\mathcal{P}$ stands for the principal part. The diagonal components $\Delta_{aa}(E_a);\Gamma_{aa}(E_a)$  are identified as the level shift (Lamb shift) and decay rate (linewidth) respectively of the synthetic and cosmological axions in absence of hybridization.

   Taking this long time limit, the amplitude equations (\ref{amp1},\ref{amp2}) become  an effective Schroedinger equation with a time \emph{independent} effective Hamiltonian
\be i\,\frac{d}{dt}\, \Bigg( \begin{array}{c}
                         A_s(t) \\
                         A_c (t)
                      \end{array}\Bigg) = \mathcal{H}_{eff} \, \Bigg( \begin{array}{c}
                         A_s(t) \\
                         A_c (t)
                      \end{array}\Bigg) \label{shcamp}\ee with

\be  \mathcal{H}_{eff} = \left(
                           \begin{array}{cc}
                             E_s+\Sigma_{ss}(E_s)  & \Sigma_{sc}(E_c) \\
                            \Sigma_{cs}(E_s)  & E_c+\Sigma_{cc}(E_d)  \\
                           \end{array}
                         \right) \equiv \left(
                                          \begin{array}{cc}
                                            \mathcal{H}_{11} & \mathcal{H}_{12} \\
                                            \mathcal{H}_{21} & \mathcal{H}_{22} \\
                                          \end{array}
                                        \right) \,.
  \label{Heffe}\ee
This effective Hamiltonian is not Hermitian, this is a manifestation that it describes the (approximate) dynamics of a \emph{quantum open system}, namely of a subset of degrees of freedom which are coupled to a continuum of other degrees of freedom whose dynamics has been ``integrated out''. Time evolution is not unitary in this subset, as is explicit from the unitarity condition (\ref{totder},\ref{unitarity}), which indicates a flow of probability from the $\ket{\phi_s},\ket{\phi_c}$ to the excited intermediate states $\ket{\big\{\kappa\big\}}$ which have been integrated out in the equations of motion.

  It proves convenient to rewrite $\mathcal{H}_{eff}$ as
  \be \mathcal{H}_{eff} = \frac{1}{2}\,\Big(E_s+\Sigma_{ss}(E_s)+E_c+\Sigma_{cc}(E_c) \Big)\,\mathbb{I}+ \frac{1}{2}\,\wD(E_s,E_c)\, \wR(E_s,E_c)\,,\label{HR}
   \ee where $\mathbb{I}$ is the $2\times 2$ identity matrix, $\wD(E_s,E_c)$ is given by
  \be \wD(E_s,E_c) = \Bigg[\Big(E_s+\Sigma_{ss}(E_s)-E_c-\Sigma_{cc}(E_c) \Big)^2+ 4\,\Sigma_{sc}(E_c)\Sigma_{cs}(E_s) \Bigg]^{1/2}\,,\label{wiD}  \ee and
   \be \wR(E_s,E_c)  = \left(
                                                                                                                         \begin{array}{cc}
                                                                                                                           \walpha(E_s,E_c) & \wbeta(E_s,E_c) \\
                                                                                                                           \wgamma(E_s,E_c) & -\walpha(E_s,E_c) \\
                                                                                                                         \end{array}
                                                                                                                       \right)\,,\label{wiR}\ee with the definitions
\bea \walpha(E_s,E_c) & = & \frac{\Big(E_s+\Sigma_{ss}(E_s)-E_c-\Sigma_{cc}(E_c) \Big)}{\wD(E_s,E_c)}\,\label{walfa}\\
\wbeta(E_s,E_c) & = & \frac{2\,\Sigma_{sc}(E_c)}{\wD(E_s,E_c)}~~; ~~ \wgamma(E_s,E_c)   =   \frac{2\,\Sigma_{cs}(E_s)}{\wD(E_s,E_c)}\,.\label{wbg}
\eea

It follows from these definitions that (suppressing the arguments $E_{s,c}$)
\be \walpha^2+\wbeta\,\wgamma =1\,,\label{wrela} \ee which implies that
\be \wR^2(E_s,E_c) = \mathbb{I} \,,\label{wr2}\ee since $Tr \,\wR =0$,  the matrix $\wR$ features eigenvalues $\pm 1$.

  Consider the eigenvalue equation (suppressing the arguments $E_{s,c}$),
  \be  \wR\, \left( \begin{array}{c}
                                                    p^\pm \\
                                                    \pm\,q^\pm
                                                  \end{array}\right) = \pm  \,  \left( \begin{array}{c}
                                                    p^\pm \\
                                                      \, q^\pm
                                                  \end{array}\right)\,,
                                         \label{eigen}\ee
 the solution of which is (up to normalization factors)
 \bea && p^+ =  \,\frac{1}{2}(1+\walpha)~~;~~ q^+ = \, \frac{\wgamma}{2} \,,\label{pqplu}\\&& p^- =  \,-\frac{\wbeta}{2}~~;~~ q^- =  \, \frac{1}{2}(1+\walpha)\,.\label{pqmin}\eea   From the relation (\ref{wrela}), we choose
 \be \walpha = \sqrt{1-\wbeta\wgamma}\,,\label{walfa2} \ee so that in absence of hybridization, $\wbeta=0;\wgamma=0$, the eigenvectors are the synthetic and cosmological axion states with eigenvalues $\pm 1$ respectively.
 The solutions (\ref{eigen}) are  eigenvectors of $\mathcal{H}_{eff}$, namely
 \be \mathcal{H}_{eff} \, \left( \begin{array}{c}
                                                    p^\pm \\
                                                      \, q^\pm
                                                  \end{array}\right) = \lambda^\pm\, \left( \begin{array}{c}
                                                    p^\pm \\
                                                      \, q^\pm
                                                  \end{array}\right) \,,\label{Heigen}\ee with eigenvalues

 \be \lambda^\pm \equiv \widetilde{\varepsilon}^{\,\pm}-i\, \frac{\widetilde{\Gamma}^{\,\pm}}{2} = \frac{1}{2}\,\Big[\Big(E_s+\Sigma_{ss}(E_s)+E_2+\Sigma_{cc}(E_c) \Big)\, \pm\,\wD(E_s,E_c) \Big]  \,,\label{lambdas}\ee where $\widetilde{\varepsilon}^{\,\pm};\widetilde{\Gamma}^{\,\pm}$ are both real and defined by equation (\ref{lambdas}). These are the   energies ($\widetilde{\varepsilon}^{\,\pm}$) and lifetimes ($ \widetilde{\Gamma}^{\,\pm}$ ) of the (quasi) normal modes resulting from the hybridization of synthetic and cosmological axions.

  The effective Hamiltonian can be diagonalized by introducing
 \be U^{-1} = \left(
                \begin{array}{cc}
                   p^+  &  p^- \\
                   q^+  & q^-  \\
                \end{array}
              \right)~~;~~ U =  \frac{1}{-p^+\,q^- + q^+ \,p^- }\, \left(
                \begin{array}{cc}
                    -q^- &   p^-  \\
                    q^+   & -p^+   \\
                \end{array}
              \right) \,,\label{Us}\ee satisfying $UU^{-1}=U^{-1}U = \mathbb{I}$, yielding
              \be U \mathcal{H}_{eff}U^{-1} =  \left(
                \begin{array}{cc}
                   \lambda^+ &  0 \\
                   0  & \lambda^- \\
                \end{array}
              \right)\,.\label{hdiag}\ee

 Let us define
 \be \Bigg( \begin{array}{c}
                         A_s(t) \\
                         A_c (t)
                      \end{array}\Bigg) = U^{-1}\,\Bigg( \begin{array}{c}
                        V_s(t) \\
                        V_c (t)
                      \end{array}\Bigg)\,, \label{AVrela} \ee the effective evolution equations for $V_{1,2}(t)$     become

 \be i\,\frac{d}{dt}\, \Bigg( \begin{array}{c}
                        V_s(t) \\
                        V_c (t)
                      \end{array}\Bigg) =   \, \Bigg( \begin{array}{c}
                        \lambda^+\, V_s(t) \\
                       \lambda^-\, V_c(t)
                      \end{array}\Bigg) \Rightarrow  \Bigg( \begin{array}{c}
                        V_s(t) \\
                        V_c (t)
                      \end{array}\Bigg) =  \left(
                                             \begin{array}{cc}
                                               e^{-i\lambda^+ t} & 0 \\
                                              0 & e^{-i\lambda^- t} \\
                                             \end{array}
                                           \right)
                         \Bigg( \begin{array}{c}
                        V_s(0) \\
                        V_c (0)
                      \end{array}\Bigg)  \,. \label{shcampV}\ee  Using the   definition (\ref{AVrela}) evaluated at $t=0$  yields the solution for the amplitudes
  \be
  \Bigg( \begin{array}{c}
                       A_s(t) \\
                        A_c(t)
                      \end{array}\Bigg) =   U^{-1}\, \left(
                                             \begin{array}{cc}
                                               e^{-i\lambda^+ t} & 0 \\
                                              0 & e^{-i\lambda^- t} \\
                                             \end{array}
                                           \right) \,U \, \Bigg( \begin{array}{c}
                         A_s(0) \\
                         A_c(0)
                      \end{array}\Bigg)\,.
 \ee With the relations (\ref{wrela},\ref{pqplu},\ref{pqmin}) it is straightforward to find that
  \be  \Bigg( \begin{array}{c}
                       A_s(t) \\
                        A_c(t)
                      \end{array}\Bigg) =  \Big[e^{-i\lambda^+ t}\,\mathbb{\wP}_+ + e^{-i\lambda^- t}\,\mathbb{\wP}_- \Big] \Bigg( \begin{array}{c}
                         A_s(0) \\
                         A_c(0)
                      \end{array}\Bigg)\,,\label{asfina} \ee  with the projector operators
                      \be \mathbb{\wP}_{\pm} = \frac{1}{2}(\mathbb{I}\pm \wR)~~;~~ \mathbb{\wP}^{\,2}_{\pm} = \mathbb{\wP}_\pm\,,\label{wprojs}\ee where $\wR$ is given by eqn. (\ref{wiR}). These project onto the quasi (because of the damping rate) normal modes. Alternatively

 \be  \Bigg( \begin{array}{c}
                       A_s(t) \\
                        A_c(t)
                      \end{array}\Bigg) =  \frac{1}{2}\Big[\Big(e^{-i\lambda^+ t}+e^{-i\lambda^- t} \Big)\,\mathbb{I} +\Big(e^{-i\lambda^+ t}-e^{-i\lambda^- t} \Big) \,\mathbb{\wR} \Big] \Bigg( \begin{array}{c}
                         A_s(0) \\
                         A_c(0)
                      \end{array}\Bigg)\,. \label{asfinalt} \ee

                      Gathering terms, we find
                      \bea A_s(t)  & = &  A^{(+)}_s(0) \,e^{-i\varepsilon^+ t}\,\,e^{-\frac{\widetilde{\Gamma}^{+}}{2}t}+ A^{(-)}_s(0) \,e^{-i\varepsilon^- t}\,\,e^{-\frac{\widetilde{\Gamma}^{-}}{2}t}\,\label{Asoft}\\
 A_c(t)  & = &  A^{(+)}_c(0) \,e^{-i\varepsilon^+ t}\,\,e^{-\frac{\widetilde{\Gamma}^{+}}{2}t}+ A^{(-)}_c(0) \,e^{-i\varepsilon^- t}\,\,e^{-\frac{\widetilde{\Gamma}^{-} }{2}t}\,\label{Acoft}\,,\eea where we used equation (\ref{lambdas}), and
 \bea  A^{(+)}_s(0) & = & \frac{1}{2}\,(1+\walpha)\,A_s(0)+\frac{\wbeta}{2}\,A_c(0)~~;~~ A^{(-)}_s(0)   =   \frac{1}{2}\,(1-\walpha)\,A_s(0)-\frac{\wbeta}{2}\,A_c(0)\,\label{Aspm}\\ A^{(+)}_c(0) & = & \frac{1}{2}\,(1-\walpha)\,A_c(0)+\frac{\wgamma}{2}\,A_s(0)~~;~~ A^{(-)}_c(0)   =   \frac{1}{2}\,(1+\walpha)\,A_c(0)-\frac{\wgamma}{2}\,A_s(0)\,\label{Acpm}
 \eea and $A_{s,c}(0)=C_{s,c}(0)$.

                      In particular, if at $t=0$ the synthetic axion is not excited, namely $A_s(0)=0$, hybridization induces their excitations for $t>0$, with amplitude in the Schroedinger picture
                      \be A_s(t) = \frac{1}{2}\,\wbeta(E_s,E_c)\,\Big(e^{-i\lambda^+ t}-e^{-i\lambda^- t} \Big)\,C_c(0) \,,\label{indusyn}\ee similarly, if at $t=0$ the cosmological axion is not excited but there is an excitation of the synthetic axion, hybridization induces cosmological excitations for $t>0$ with amplitude
                      \be A_c(t) = \frac{1}{2}\,\wgamma(E_s,E_c)\,\Big(e^{-i\lambda^+ t}-e^{-i\lambda^- t} \Big)\,C_s(0) \,.\label{inducos}\ee

                       This is one of the main results of this study: excitations of synthetic axions are induced by the presence of cosmological axions or viceversa as a consequence of hybridization.

                      Inserting the above results in the equation for the coefficients $C_{\kappa}(t)$ of the intermediate states, eqn. (\ref{cmoft}) we find the amplitude for the two-photon state
  \bea C_{\kappa} (t) & = & \Bigg[ T_{\kappa s}\,A^{(+)}_s(0)+ T_{\kappa c}\,A^{(+)}_c(0)\Bigg] \Bigg[ \frac{1-e^{i(E_\kappa-\varepsilon^+ + i  \frac{\widetilde{\Gamma}^+}{2})t}}{E_\kappa-\varepsilon^+ + i  \frac{\widetilde{\Gamma}^+}{2}}   \Bigg]\nonumber \\ & + &  \Bigg[ T_{\kappa s}\,A^{(-)}_s(0)+ T_{\kappa c}\,A^{(-)}_c(0)\Bigg] \Bigg[ \frac{1-e^{i(E_\kappa-\varepsilon^- + i  \frac{\widetilde{\Gamma}^-}{2})t}}{E_\kappa-\varepsilon^- + i  \frac{\widetilde{\Gamma}^-}{2}}   \Bigg] \,.\label{cikapas} \eea

  The probabilities for synthetic and cosmological axions feature interference terms,
  \be |C_s(t)|^2 = |A_s^+(0)|^2\,e^{-\widetilde{\Gamma}^+ t}+ |A_s^-(0)|^2\,e^{-\widetilde{\Gamma}^- t}+ 2\mathrm{Re} \Big[A_s^+(0)(A_s^-(0))^*\,e^{-i(\varepsilon^+ -\varepsilon^-)t}\,e^{-\frac{1}{2}(\widetilde{\Gamma}^++\widetilde{\Gamma}^-)t}\Big]\,,\label{interfcs}\ee with a similar expression for $|C_c(t)|^2$ with $A^{\pm}_s(0) \rightarrow A^{\pm}_c(0)$. The oscillatory interference terms are a consequence of hybridization, as can be seen explicitly since $A^{-}_s(0), A^{+}_c(0)$ vanish as $\widetilde{\beta},\widetilde{\gamma} \rightarrow 0$,  and are a manifestation of quantum beats. As a consequence of the unitarity relation (\ref{unitarity}) these quantum beats are also manifest in the probabilities $|C_{\kappa}(t)|^2$ of the intermediate states, and will be explicitly shown below.

  These  results  are general (in the Markov approximation)  and complement and extend those of references\cite{garraway,plenio} by including the full contribution from the self-energy corrections, not just the imaginary parts.

   \vspace{1mm}

  \textbf{Important checks:} The validity of the description of hybridization can be confirmed with the following checks.

  \begin{itemize}
  \item{Setting $C_c(0)=0;g_c=0$ the set of equations (\ref{eqnc1}-\ref{eqncm}) simplifies to those obtained from the Weisskopf-Wigner  description of spontaneous emission in a two levels  atom\cite{zubairy}, (similarly if $C_s(0)=0;g_s=0$}).

  \item{Neglecting the real part of the self-energies, namely setting the Lamb-shifts $\Delta_{ab}=0$,  recover the results of ref.\cite{garraway,plenio}}. In particular, the ``mode'' described by the $C_{\kappa}$ is equivalent to the ``pseudomode'' in ref.\cite{garraway}.

  \end{itemize}

  The equivalence to the case of spontaneous emission in a two-level atom, obtained by setting $g_c=0;C_c(0)=0$ will prove important because the results obtained in the general
  case of hybridization will also apply to the case of just the synthetic emergent axion decaying into two photons. This result, a byproduct of the more general treatment with hybridization will be used in section (\ref{subsec:cmaxion}) to argue that two-photon (HBT) correlations may also be an important experimental probe of axion collective excitations in topological insulators and/or Weyl semimetals.

  \section{Synthetic and cosmological axion hybridization:}\label{subsec:hybrid}

  In this section we   apply the general results of the previous section to the hybridization of the synthetic and cosmological axions for which  the interaction Hamiltonian to leading order in the couplings is given by eqn. (\ref{HI}).

  In interaction picture (see appendix (\ref{app:hamcs})) $\vec{E}(\vx,t) = -\dot{\vec{A}}(\vx,t)~;~\vec{B}(\vx,t) = \vec{\nabla}\times \vec{A}(\vx,t)$ and $\vec{A}(\vx,t)$ is quantized in Coulomb gauge with the expansion (\ref{vecpot}) with real (linear) polarization vectors. The pseduscalar fields $\phi_{s,c}(\vx,t)$ in interaction picture are expanded as in eqn. (\ref{scalarquant2}).

    We need the transition matrix elements
  \be T_{s\kappa} = \langle 1_{\vk,s}|H_I(0)|\kappa\rangle ~~; ~~ T_{c\kappa} = \langle 1_{\vk,c}|H_I(0)|\kappa\rangle\,,\label{mtxT}\ee where $\ket{1_{\vk,s,c}}$ are single
  particle states of spatial momentum $\vk$  of the synthetic and cosmic axions and the vacuum for the electromagnetic fields. Evidently the states $\ket{\kappa}$ are two-photon states of the
  form $\ket{\kappa}=\ket{1_{\vk_1,\lambda_1},1_{\vk_2,\lambda_2}}$ with $\vec{k}_2=\vec{k}-\vec{k}_1$, and energy
  \be E_{\kappa} \equiv E_{2\gamma}= k_1+k_2 \,. \label{Ekapa}\ee

   Carrying out the spatial integral, we find the matrix elements for $a=s,c$
  \be T_{a\kappa}\equiv \langle 1_{\vk,a}|H_I(0)|1_{\vk_1,\lambda_1},1_{\vk_2,\lambda_2}\rangle= \frac{g_a\, {k_1k_2}}{2\sqrt{2\,E_{a}(k)V}}\,  \,\Bigg[\vec{\epsilon}_{\lambda_1}(\vk_1) \cdot \Big(\hat{\vk}_2 \times \vec{\epsilon}_{\lambda_2}(\vk_2)\Big)+ \vec{\epsilon}_{\lambda_2}(\vk_2)\cdot \Big(\hat{\vk}_1 \times \vec{\epsilon}_{\lambda_1}(\vk_1)\Big) \Bigg]_{\vec{k}_2=\vec{k}-\vec{k}_1} \,.\label{Tmx}   \ee With  sums replaced by integrals as
  \be \sum_{\kappa} = \sum_{\vk_1,\vk_2}\sum_{\lambda_1,\lambda_2} ~~;~~ \frac{1}{V}\sum_{\vk_1} \rightarrow \int \frac{d^3k_1}{(2\pi)^3}\,, \ee the spectral density (\ref{rhosd}) becomes
\be \rho_{ab}(k_0;k) = \frac{g_a g_b}{4\sqrt{E_a(k)E_b(k)}}\,\int\frac{d^3k_1}{(2\pi)^3}\frac{1}{k_1|\vec{k}-\vec{k}_1|}\,\Big[k_1|\vec{k}-\vec{k}_1|-\vk_1\cdot(\vk-\vk_1) \Big]^2 \,\delta(k_0-k_1-|\vec{k}-\vec{k}_1|) \,,\label{rhoabfin}\ee
The details of its calculation are provided in appendix (\ref{app:spec}), with the result
\be \rho_{ab}(k_0;k) = \frac{g_a g_b}{32\pi^2\sqrt{E_a(k)E_b(k)}}\Big(k^2_0-k^2 \Big)^2\,\Theta(k_0-k)\,, \label{rhoabfinal} \ee yielding
for the real and imaginary parts of the self-energy matrix
\bea \Delta_{ss}(E_s) & = & \frac{g^2_s}{32\pi^2 E_s(k)}\int^{\Lambda}_k \mathcal{P} \Big[ \frac{(k^2_0-k^2)^2}{E_s-k_0} \Big]  ~~;~~ \Gamma_{ss}(E_s) =  \frac{g^2_s\,(E^2_s(k)-k^2)^2}{16\pi E_s(k)}\,, \nonumber\\ \Delta_{cc}(E_c) & = & \frac{g^2_c}{32\pi^2 E_c(k)}\int^{\Lambda}_k \mathcal{P} \Big[ \frac{(k^2_0-k^2)^2}{E_c-k_0} \Big]  ~~;~~ \Gamma_{cc}(E_c) =  \frac{g^2_c\,m^2_c }{16\pi E_c(k)}\,, \nonumber\\ \Delta_{sc}(E_c) & = & \frac{g_s g_c}{32\pi^2 \sqrt{E_s(k)E_c(k)}}\int^{\Lambda}_k \mathcal{P} \Big[ \frac{(k^2_0-k^2)^2}{E_c-k_0} \Big]  ~~;~~ \Gamma_{sc}(E_c) = \frac{g_s g_c\,m^2_c}{16\pi \sqrt{E_s(k)E_c(k)}} \,, \nonumber\\ \Delta_{cs}(E_s) & = & \frac{g_s g_c}{32\pi^2 \sqrt{E_s(k)E_c(k)}}\int^{\Lambda}_k \mathcal{P} \Big[ \frac{(k^2_0-k^2)^2}{E_s-k_0} \Big]  ~~;~~ \Gamma_{cs}(E_s) = \frac{g_s g_c\,(E^2_s(k)-k^2)^2}{16\pi \sqrt{E_s(k)E_c(k)}} \,, \label{reimsigs}\eea where we have introduced an upper cutoff $\Lambda$ in the integrals for the real parts (Lamb-shifts $\Delta_{ab}$) to highlight that all these
contributions are ultraviolet divergent in the $\Lambda \rightarrow \infty$ limit requiring renormalization, an aspect that will be discussed in section (\ref{sec:discussion}).

We are particularly interested in the amplitude of the two-photon state (\ref{cikapas}), which becomes
\be C_{2\gamma}(\vec{k}_1,\lambda_1;\vec{k_2},\lambda_2;t)   =   \Bigg\{\mathcal{A}^+\Bigg[ \frac{1-e^{i(E_{2\gamma}-\varepsilon^+ + i  \frac{\widetilde{\Gamma}^+}{2})t}}{E_{2\gamma}-\varepsilon^+ + i  \frac{\widetilde{\Gamma}^+}{2}}   \Bigg]  +   \mathcal{A}^- \Bigg[ \frac{1-e^{i(E_{2\gamma}-\varepsilon^- + i  \frac{\widetilde{\Gamma}^-}{2})t}}{E_{2\gamma}-\varepsilon^- + i  \frac{\widetilde{\Gamma}^-}{2}}   \Bigg] \Bigg\} \, \mathcal{C}(\vec{k}_1,\lambda_1;\vec{k_2},\lambda_2) \label{c2gama}   \,,  \ee where $E_{2\gamma} = k_1+|\vec{k}-\vec{k}_1|$, and  the amplitudes $\mathcal{A}^{\pm}$ are given by
\bea \mathcal{A}^+ & = &  \frac{  {k_1|\vec{k}-\vec{k}_1|}}{\sqrt{8 V}}\,\Bigg[\frac{g_s\,A^{(+)}_s(0)}{\sqrt{E_s(k)}} + \frac{g_c\,A^{(+)}_c(0)}{\sqrt{E_c(k)}}\Bigg] \,\label{amplu} \\
 \mathcal{A}^- & = & \frac{  {k_1|\vec{k}-\vec{k}_1|}}{\sqrt{8 V}}\,\Bigg[\frac{g_s\,A^{(-)}_s(0)}{\sqrt{E_s(k)}} + \frac{g_c\,A^{(-)}_c(0)}{\sqrt{E_c(k)}}\Bigg] \,, \label{amin}\eea
 with   $A^{\pm}_{s,c}(0)$   given by equations (\ref{Aspm},\ref{Acpm}). The function
\be \mathcal{C}(\vec{k}_1,\lambda_1;\vec{k}_2,\lambda_2)  =    \,\Big[\vec{\epsilon}_{\lambda_1}(\vk_1) \cdot \Big(\hat{\vk}_2 \, \times \, \vec{\epsilon}_{\lambda_2}(\vk_2)\Big)+ \vec{\epsilon}_{\lambda_2}(\vk_2)\cdot \Big(\hat{\vk}_1 \times \vec{\epsilon}_{\lambda_1}(\vk_1)\Big) \Big]~~;~~\vec{k}_2=\vec{k}-\vec{k}_1\,,\label{Cpola} \ee contains all the polarization information of the photon pair, depends only on the direction of the wavevectors,  and is symmetric under the exchange $\vec{k}_1,\lambda_1 \Leftrightarrow \vec{k}_2,\lambda_2$, which implies that
\be C_{2\gamma}(\vec{k}_1,\lambda_1;\vec{k}_2,\lambda_2;t)= C_{2\gamma}(\vec{k}_2;\lambda_2,\vec{k}_1,\lambda_1;t) \,.  \label{symm}\ee

The  two photon state
\be \ket{\psi_{2\gamma}(t)} = \sum_{\vec{k}_1,\lambda_1,\lambda_2} C_{2\gamma}(\vec{k}_1,\lambda_1;\vec{k_2},\lambda_2;t)\,\ket{1^\gamma_{\vec{k}_1,\lambda_1};1^\gamma_{\vec{k}_2,\lambda_2}}~;~\vec{k}_2=\vec{k}-\vec{k}_1\,,\label{twogamast}\ee is entangled both in momentum and polarization with two noteworthy aspects:

\vspace{1mm}

\textbf{i:)} the resonant denominators in the amplitudes (\ref{c2gama}), entail that the probability $|C_{2\gamma}(\vec{k}_1,\lambda_1,\vec{k_2},\lambda_2;t) |^2 $ of finding a photon pair with momenta $\vec{k}_1,\vec{k}_2=\vec{k}-\vec{k}_1$ features two peaks, at the values $k_1+|\vec{k}-\vec{k}_1| \simeq \varepsilon^{\pm}$, which are the kinematic regions of energy conserving transitions. If $|\varepsilon^+-\varepsilon^- | \lesssim \Gamma^+ +\Gamma^-$ the two peaks ``blur'' into one, this is the (unlikely) case when the synthetic and cosmological axions are nearly degenerate (see discussion below).

\vspace{1mm}

\textbf{ii:)} To highlight polarization entanglement in a clear manner, let us consider the case $\vk =0$, namely $\vec{k}_2 = - \vec{k}_1$ corresponding to back-to-back photons. In this case the coefficient (\ref{Cpola}) becomes
\be \mathcal{C}(\vec{k}_1,\lambda_1;-\vec{k}_1,\lambda_2)  =  \Big[\vec{\epsilon}_{\lambda_1}(\vk_1) \cdot \Big(-\hat{\vk}_1 \times \vec{\epsilon}_{\lambda_2}(-\vk_1)\Big)+ \vec{\epsilon}_{\lambda_2}(-\vk_1)\cdot \Big(\hat{\vk}_1 \times \vec{\epsilon}_{\lambda_1}(\vk_1)\Big) \Big]\,, \label{cpolakzero}\ee using the properties (\ref{polas}) of the polarization vectors, we find that if $\lambda_1 =1$ then $\lambda_2=2$ for the amplitude to be non-vanishing, and viceversa, implying that the photons are emitted back-to-back and with perpendicular polarizations. In other words, the emitted photons feature   telltale correlations both in momenta and in polarization. The polarization dependence of the coefficient $\mathcal{C}(\vec{k}_1,\lambda_1;\vec{k}_2,\lambda_2)$ in eqn. (\ref{Cpola}) is a distinct manifestation of the pseudoscalar coupling to electromagnetism via the Chern-Simons term.

\subsection{HBT correlations (second order coherence):}\label{subsec:hbt}
The characteristics of the two photon state may be probed with intensity interferometry as per Glauber's theory of photodetection\cite{glauber} with the
Hanbury-Brown Twiss\cite{hbt} (HBT) second order coherence for coincident photo detection
\be G^{(2)}(\vec{x}_1,t;\vec{x}_2,t) = \bra{\psi(t)}E^{-}(\vec{x}_2,t)E^{-}(\vec{x}_1,t)E^{+}(\vec{x}_1,t)E^{+}(\vec{x}_2,t)\ket{\psi(t)} \,,\label{G2}\ee (we suppressed the vector labels) where
\bea \vec{E}^{-}(\vec{x},t)  & = & \frac{-i}{\sqrt{V}}\sum_{\vk,\lambda=1,2} \sqrt{\frac{k}{2}}\, \vec{\epsilon}_{\lambda}(\vk)\,a^\dagger_{\vk,\lambda}\,e^{i(\vk\cdot \vx-kt)}\,,\label{Emin}\\
\vec{E}^{+}(\vec{x},t)  & = & \frac{i}{\sqrt{V}}\sum_{\vk,\lambda=1,2} \sqrt{\frac{k}{2}}\, \vec{\epsilon}_{\lambda}(\vk)\,a_{\vk,\lambda}\,e^{-i(\vk\cdot \vx-kt)}\,. \label{Eplus}\eea
As discussed above, the two photon state features correlations both in momentum and polarization. However,  the operators (\ref{Emin},\ref{Eplus}) are   sums of both observables, hence, the spatio-temporal second order coherence (\ref{G2}) probes momentum and polarization entanglement only indirectly. Following ref.\cite{mandl} we introduce instead the momentum and polarization resolved operators
\bea \mathcal{O}^{-}_{\vec{p},s}(t) & = &  {\sqrt{\frac{1}{V}}}\int   \vec{\epsilon}_s(\vec{p})\cdot \vec{E}^{-}(\vx,t) \,e^{-i \vec{p}\cdot \vx}  \, d^3 x \,,\label{Omin}\\\mathcal{O}^{+}_{\vec{p},s}(t) & = &  {\sqrt{\frac{1}{V}}}\int  \vec{\epsilon}_s(\vec{p})\cdot \vec{E}^{+}(\vx,t) \,e^{i \vec{p}\cdot \vx} \, d^3 x\,,\label{Oplus} \eea which are directly related to the photon annihilation and creation operators as
\be \mathcal{O}^{-}_{\vec{p},s}(t) = \sqrt{\frac{p}{2}}\,a_{\vec{p},s}\,e^{-ikt}~~;~~ \mathcal{O}^{+}_{\vec{p},s}(t) = \sqrt{\frac{p}{2}}\,a^\dagger_{\vec{p},s}\,e^{ikt}\,. \label{Oarel}\ee Instead of the spatio-temporal (HBT) correlation (\ref{G2}), and following ref.\cite{mandl} we obtain the momentum and polarization resolved cross-correlation (second order cross-spectral density)\footnote{See chapter 12.5.2 in ref.\cite{mandl}.}
\be \widetilde{G}^{(2)}(\vec{p}_1,s_1;\vec{p}_2,s_2;t) =  \bra{\psi(t)}a^{\dagger}_{\vec{p}_1,s_1}a^{\dagger}_{\vec{p}_2,s_2}a_{\vec{p}_2,s_2}a_{\vec{p}_1,s_1}\ket{\psi(t)} \,,\label{tilG2}\ee  from which the full spatio-temporal second order coherence (\ref{G2})  may be obtained.
Using the symmetry (\ref{symm}) we find
\be \widetilde{G}^{(2)}(\vec{p}_1,s_1;\vec{p}_2,s_2;t) = 4\,\big|C_{2\gamma}(\vec{p}_1,s_1;\vec{p}_2,s_2;t)\big|^2\,\delta_{\vec{p}_1+\vec{p}_2,\vec{k}}\,.\label{tilGfin} \ee

It is straightforward to obtain also the first order cross-correlation
\be  \widetilde{G}^{(1)}(\vec{p}_1,s_1;\vec{p}_2,s_2;t) =  \bra{\psi(t)}a^{\dagger}_{\vec{p}_1,s_1}a_{\vec{p}_2,s_2} \ket{\psi(t)} \,,\label{tilG1}\ee  however, the result is neither very illuminating nor useful as a probe of polarization  because it entails the sum over one of the polarizations and   averages over the polarization information. However, for completeness this correlation function is obtained in appendix (\ref{app:1stcorr}).

Therefore, we propose the momentum and polarization resolved two-photon (HBT) cross-correlation (\ref{tilG2}) as a possible experimental probe of both momentum and polarization entanglement with the distinct polarization correlations determined by the pseudoscalar coupling, a telltale of axion electrodynamics encoded in the coefficient (\ref{Cpola}).

Armed with the general result (\ref{c2gama}) we find
\be  \widetilde{G}^{(2)}(\vec{p}_1,s_1;\vec{p}_2,s_2;t) = 4\,\Big\{|\mathcal{A}^+|^2\,D^+(\vec{p}_1;t)+ |\mathcal{A}^-|^2\,D^-(\vec{p}_1;t)+\big(\mathcal{A}^+\big)^*\mathcal{A}^-\,I[\vec{p}_1;t] + c.c.  \Big\} \, \mathcal{C}^2(\vec{p}_1,s_1;\vec{p}_2,s_2) \label{tilG2fin}\ee
were $\vec{p}_2= \vec{k}-\vec{p}_1$. The direct term $D^{+}$  is  given by
\be D^+(\vec{p}_1;t) = \frac{\Bigg[ 1-\Big(e^{i(E_{2\gamma}-\varepsilon^+)t} + e^{-i(E_{2\gamma}-\varepsilon^+)t}\Big)\,e^{-\frac{\widetilde{\Gamma}^+}{2}t}+ e^{- {\widetilde{\Gamma}^+} t}      \Bigg]}{(E_{2\gamma}-\varepsilon^+)^2 + \Big(\frac{ \widetilde{\Gamma}^+ }{2}\Big)^2}\,,\label{Dplus}\ee with a similar expression for $D^-(\vec{p}_1;t)$ with the substitution $\varepsilon^+\, , \,  \widetilde{\Gamma}^+ \rightarrow \varepsilon^-\,,\, \widetilde{\Gamma}^-$. The interference term is given by
\be I[\vec{p}_1;t]= \frac{1-e^{-i(E_{2\gamma}-\varepsilon^+)t}\, e^{-\frac{\widetilde{\Gamma}^+}{2}t}-e^{i(E_{2\gamma}-\varepsilon^-)t}\, e^{-\frac{\widetilde{\Gamma}^-}{2}t}+ e^{i(\varepsilon^+-\varepsilon^-)t}\,e^{-\frac{(\widetilde{\Gamma}^++\widetilde{\Gamma}^-)}{2}t}}{\big( E_{2\gamma}-\varepsilon^+ - i  \frac{\widetilde{\Gamma}^+}{2}\big)\, \big( E_{2\gamma}-\varepsilon^- + i  \frac{\widetilde{\Gamma}^-}{2}\big)}\,.  \label{interf}\ee

The direct and interference terms are dominated by the region of $E_{2\gamma}$ near the resonances, namely $E_{2\gamma} \simeq \varepsilon^{\pm}$, and the interference term displays   the quantum beats with typical frequency $\varepsilon^{+}-\varepsilon^{-}$. The dominant part of the direct and interference terms can be made more explicit  by ``filtering'' the cross correlation in frequency near the resonances.

Momentum or frequency resolved (filtered) (HBT) correlations including cross-correlations of the form (\ref{tilG2})  have been experimentally studied as
a ``colored'' (HBT) in ref.\cite{hbtcolor} where the source of light is a polariton condensate. This experiment demonstrates the feasibility of frequency resolution in intensity interferometry, suggesting the possible experimental   feasibility of the  momentum and polarization resolved   (HBT) cross-correlation (\ref{tilGfin})  to probe two-photon correlated states from the hybridization of synthetic and cosmological axions.

The contributions to the (HBT) cross correlation $\widetilde{G}^{(2)}$ while not being monochromatic, are dominated by the respective  resonant denominators, however photodetectors feature a finite resolution,  therefore following the filtering procedure of ref.\cite{hbtcolor} we integrate this correlation in frequency  (or $E_{2\gamma}$), but not in direction so as to maintain the polarization dependence,  within a small band of width $\xi$ (determined by the photodetector resolution) around the resonances
$\xi <  |E_{2\gamma}-\varepsilon|$ with $ \Gamma^{\pm} \ll \xi \ll E_{2\gamma}$. Assuming very narrow widths $\Gamma^{\pm}$ for weak couplings, we extend the integration region to infinity and carry out the integration by contour integration in the complex frequency plane\footnote{See a similar treatment in chapter 6.3 of reference\cite{zubairy}.} as in refs.\cite{zubairy,scully}.

For terms containing exponentials of the form $ e^{i(E_{2\gamma}-\varepsilon^+)t} $  ($ e^{-i(E_{2\gamma}-\varepsilon^+)t} $ ) the contour is closed in the upper (lower) half frequency plane yielding for the integration of the direct and interference terms in $\widetilde{G}^{(2)}$
\be D^+(\vec{p}_1;t) \simeq 2 \pi \frac{1-e^{-\Gamma^+ t}}{\Gamma^+}~~;~~ D^-(\vec{p}_1;t) \simeq 2 \pi \frac{1-e^{-\Gamma^- t}}{\Gamma^-}\,,\label{intdirs}\ee

\be  I[\vec{p}_1;t] \simeq  2\pi\,i \Bigg[ \frac{1- e^{i(\varepsilon^+-\varepsilon^-)t}\,e^{- {(\widetilde{\Gamma}^++\widetilde{\Gamma}^-)} \frac{t}{2}}}{ \varepsilon^+-\varepsilon^- + \frac{i}{2} \, \big( {\widetilde{\Gamma}^++\widetilde{\Gamma}^-}    \big)}\Bigg] \,.  \label{interbeats}  \ee  Gathering all the terms, upon filtering in a narrow band of frequencies near the resonances we find the final form of the second order cross correlation
\bea  &&  \widetilde{G}^{(2)}(\vec{p}_1,s_1;\vec{p}_2,s_2;t)  \propto  \Bigg\{\frac{|\mathcal{A}^+|^2}{\Gamma^+}(1-e^{-\Gamma^+ t})+\frac{|\mathcal{A}^-|^2}{\Gamma^-}(1-e^{-\Gamma^- t})\nonumber \\ &  + &  i \,\big(\mathcal{A}^+\big)^*\mathcal{A}^-\, \Bigg[\frac{1- e^{i(\varepsilon^+-\varepsilon^-)t}\,e^{- {(\widetilde{\Gamma}^++\widetilde{\Gamma}^-)} \frac{t}{2}}}{ \varepsilon^+-\varepsilon^- + \frac{i}{2} \, \big( {\widetilde{\Gamma}^++\widetilde{\Gamma}^-}    \big)}\Bigg]+ c.c \Bigg\} \,\mathcal{C}^2(\vec{p}_1,s_1;\vec{p}_2,s_2)\,. \label{tilG2final}  \eea  The polarization function $\mathcal{C}(\vec{p}_1,s_1;\vec{p}_2,s_2)$ only depends on the directions of the vectors which have not been integrated over so that polarization information is maintained after frequency integration within the photodetector bandwidth.

The last term inside the brackets in the final result (\ref{tilG2final}), arising from the interference contribution $I[\vec{p}_1;t]$ displays the quantum beats.

A noteworthy aspect of the above results is that, the ratios $ {|\mathcal{A}^\pm|^2}/{\Gamma^\pm}$ are of $\mathcal{O}(1)$ in the couplings, whereas the interference term is quadratic in the couplings, however, it is enhanced in the (unlikely) nearly degenerate case when $|\varepsilon^+-\varepsilon^-|$ is of order $g^2_s,g^2_c, g_sg_c$.

 \textbf{Non-degenerate limit:} The results (\ref{c2gama}) and (\ref{tilG2final}) are general and   valid    for arbitrary small couplings $g_s,g_c$ and masses $m_s,m_c$. However, it is very unlikely that the masses of the synthetic and cosmological axions be nearly degenerate and of the same order as the self-energy corrections $\Sigma_{ab},a,b=s,c$  in the expressions (\ref{HR},\ref{wiD}). Neglecting the possibility of a (very unlikely) coincidence of the masses, we consider that $|E_s-E_c| \gg \Sigma_{ab}$ since the self-energy matrix is of quadratic order in the small couplings,  referring to this case as the \emph{non-degenerate limit}. With this assumption the expressions for  (\ref{wiD}-\ref{wbg}) simplify  to
  \bea  \wD(E_s,E_c) & \simeq &   |E_s+\Sigma_{ss}(E_s)-E_c-\Sigma_{cc}(E_c)| + \mathcal{O}(g^2_s g^2_c) \,\label{wiDapp}\\
   \walpha(E_s,E_c) & \simeq  & 1 +\mathcal{O}(g^2_sg^2_c) ~~;~~
\wbeta(E_s,E_c)   \simeq   \mathcal{O}(g_s g_c)  ~~; ~~ \wgamma(E_s,E_c)       \simeq   \mathcal{O}(g_s g_c)\,.\label{wbgapp} \eea In the following analysis we will consider that $E_s > E_c$, corresponding to the choice $\walpha(E_s,E_c) \simeq 1$, in this non-degenerate limit,  the case $E_c > E_s$ can be obtained from straightforward replacements.

In this limit the (quasi) normal mode eigenvalues  are given by

\bea \lambda^+  & = &  E_s + \Sigma_{ss}(E_s) + \mathcal{O}(g^2_s g^2_c)\,,\label{lamplu}\\
\lambda^-  & = &  E_c + \Sigma_{cc}(E_c)+ \mathcal{O}(g^2_s g^2_c) \,,\label{lamin} \eea where the real and imaginary parts of the self energies are given by equation (\ref{RIsig}). The real part of the diagonal matrix elements of the self-energy correspond to renormalizations of $E_{s,c}$ (Lamb shifts), and are absorbed into the renormalized energies
\be E^{r}_{s} = E_{s}+\Delta_{ss}(E_s);E^{r}_{c} = E_{c}+\Delta_{cc}(E_c)\,,\label{ERs}\ee
 therefore, in the non-degenerate limit we find the (quasi) normal mode frequencies
\bea \lambda^+  & = &  E^r_s -i \frac{\Gamma_{ss}}{2} ~~;~~ \Gamma_{ss} = 2\pi \rho_{ss}(E_s) \,,\label{lamplufin}\\
 \lambda^-  & = &  E^r_c -i \frac{\Gamma_{cc}}{2} ~~;~~ \Gamma_{cc} = 2\pi \rho_{cc}(E_c)  \,,\label{laminfin} \eea where the diagonal matrix elements of the spectral density $\rho_{ab}$ and decay widths  can be read off equations (\ref{reimsigs}).

Since the matrix elements $T_{a\kappa}$ given by eqn. (\ref{Tmx}) are of order $g_a$, it follows that    we can replace the amplitudes (\ref{Aspm},\ref{Acpm}) in equation (\ref{cikapas})    by their leading order in the couplings, namely
 \bea  A^{(+)}_s(0) & \simeq &  A_s(0)=C_s(0) ~~;~~ A^{(-)}_s(0)   \simeq  0  \,\label{Aspmnd}\\ A^{(+)}_c(0) & \simeq  & 0 ~~;~~ A^{(-)}_c(0)   \simeq   A_c(0)=C_c(0)  \,.\label{Acpmnd}
 \eea In this approximation the two-photon amplitude (\ref{c2gama}) becomes

\bea C_{2\gamma}(\vec{p}_1,s_1;\vec{p}_2,s_2;t) & = & \Bigg\{  \frac{g_s\,C_s(0)}{\sqrt{E_s(k)}}    \Bigg[ \frac{1-e^{i(E_{2\gamma}-E^r_{s}+ i  \frac{\Gamma_{ss}}{2})t}}{E_{2\gamma}-E^r_{s} + i  \frac{\Gamma_{ss}}{2}}   \Bigg] +    \frac{g_c\,C_c(0)}{\sqrt{E_c(k)}} \Bigg[ \frac{1-e^{i(E_{2\gamma}-E^r_c+ i  \frac{\Gamma_{cc}}{2})t}}{E_{2\gamma}-E^r_c + i  \frac{\Gamma_{cc} }{2}}   \Bigg] \Bigg\}\nonumber\\ &  \times & \mathcal{C}(\vec{p}_1,s_1;\vec{p}_2,s_2)\,\delta_{\vec{p}_2,\vec{k}-\vec{p}_1} ~~;~~ E_{2\gamma} = p_1+ p_2  \label{c2gamand}  \,.  \eea We note  that in keeping only the leading terms in the couplings, we have neglected the ``induced'' amplitudes (\ref{indusyn},\ref{inducos}) which are  a consequence of hybridization and suppressed by the product $g_s g_c$ with respect to the leading order terms in the non-degenerate limit. However, if either of the initial amplitudes $C_{s,c}(0)$ vanish, these subleading ``induced'' terms must be kept, since these will be the source of interference and quantum beats.

In this non-degenerate case, and assuming   initial amplitudes $C_s(0)\neq 0;C_c(0)\neq 0$,   the frequency filtered second order cross correlation(\ref{tilG2final}) becomes

\bea  &&  \widetilde{G}^{(2)}(\vec{p}_1,s_1;\vec{p}_2,s_2;t)  \propto  \Bigg\{\frac{g^2_s \,|C_s(0)|^2}{E_s(k)\Gamma_{ss}}(1-e^{-\Gamma_{ss} t})+\frac{g^2_c\,|C_c(0)|^2}{E_c(k)\Gamma_{cc}}(1-e^{-\Gamma_{cc} t})\nonumber \\ &  + &  i \,\frac{g_s g_c C^*_s(0)C_c(0)}{\sqrt{E_s(k)E_c(k)}}\, \Bigg[\frac{1- e^{i(E^r_s-E^r_c)t}\,e^{- {( {\Gamma}_{ss}+ {\Gamma}_{cc})} \frac{t}{2}}}{ E^r_s-E^r_c+ \frac{i}{2} \, \big( { {\Gamma}_{ss}+ {\Gamma}_{cc}}    \big)}\Bigg]+ c.c \Bigg\} \,\mathcal{C}^2(\vec{p}_1,s_1;\vec{p}_2,s_2)\,, \label{tilG2finnd}  \eea with $\mathcal{C}(\vec{p}_1,s_1;\vec{p}_2,s_2)$ being the polarization function (\ref{Cpola}).  The quantum beats are explicit in the third, interference term with the beat frequency being the difference between the (renormalized) synthetic and cosmological axion energies.
 This is one of the main results of our study.

\section{Detecting synthetic axions with (HBT):}\label{subsec:cmaxion}
The discussion above focused on understanding  the consequences  of hybridization between synthetic and cosmological axions, such as  momentum and polarization entanglement and in particular quantum beats as a manifestation of interference. However,  momentum and polarization entanglement are characteristics of the process of axion decay that apply individually to each  species independently of their hybridization. Kinematic entanglement is a consequence of energy-momentum conservation, and  the particular pattern of polarization entanglement is solely a consequence of the Chern-Simons coupling to electromagnetism, which is common to both species of axions. This is explicit in the expression (\ref{tilG2finnd}) where the polarization dependent function $\mathcal{C}$ multiplies all the terms.

The feebleness of the coupling of cosmic axions to electromagnetism\cite{marsh,sikivie1,graham,irastorza}  may result in a quantum beat signal, the third term in equation (\ref{tilG2finnd}) (valid in the non-degenerate case), to be  too small to be detected.

 However, even in this case, the analysis above unambiguously suggests that momentum and polarization resolved (HBT) second order coherence may be a useful experimental probe to detect \emph{synthetic axions  independently}.

The case of synthetic axions only can be obtained from the non-degenerate limit straightforwardly by setting $g_c=0$, yielding $\walpha(E_s,E_c) =1,\wbeta=0,\wgamma=0$ and $\varepsilon^+ = E^r_s$. Therefore, the two-photon amplitude (\ref{c2gamand}) from \emph{synthetic axion} decay  becomes
\be  C_{2\gamma}(\vec{p}_1,s_1;\vec{p}_2,s_2;t)  =   \frac{g_s\,C_s(0)}{\sqrt{E_s(k)}}    \Bigg[ \frac{1-e^{i(E_{2\gamma}-E^r_{s}+ i  \frac{\Gamma_{ss}}{2})t}}{E_{2\gamma}-E^r_{s} + i  \frac{\Gamma_{ss}}{2}}   \Bigg] \,\mathcal{C}(\vec{p}_1,s_1;\vec{p}_2,s_2)\,\delta_{\vec{p}_2,\vec{k}-\vec{p}_1} ~~;~~ E_{2\gamma} = p_1+ p_2  \label{c2gamasyn}  \,,  \ee and the frequency-filtered second order cross correlation (\ref{tilG2finnd}) becomes
\be \widetilde{G}^{(2)}(\vec{p}_1,s_1;\vec{p}_2,s_2;t)  \propto    \frac{2\pi g^2_s\,|C_s(0)|^2}{E_s\Gamma_{ss}}(1-e^{-\Gamma_{ss} t})\,\mathcal{C}^2(\vec{p}_1,s_1;\vec{p}_2,s_2)\,.\label{tilG2syn}\ee

The distinct pattern of polarization entanglement, determined by the function $\mathcal{C}$ in (\ref{tilG2syn}) and momentum entanglement are     features  of the two-photon state from synthetic axion decay independently of the hybridization with the cosmological axion.  Therefore our argument above suggests that coincident two photon photodetection with momentum and polarization resolution via the   (HBT) second order coherence or cross correlation is a robust tool to seek signatures of synthetic axions.  (HBT) second order coherence may provide a useful diagnostic tool complementing   the proposal in ref.\cite{olivia} for multiphoton detection as a probe of axionic collective excitations in topological insulators or Weyl semimetals.

\subsection{Similarities and differences with parametric down conversion:}\label{subsec:pdc}
The results in the previous sections bear a striking similarity to the phenomenon of parametric down conversion\cite{pdc1,pdc2,pdc3}. In this process   an incoming
beam of photons from a laser (pump) propagate in a non-linear medium and split into two lower frequency signal and idler photons through a non-linear second order susceptibility tensor $\chi_{ijk}$. The interaction Hamiltonian is modeled with a cubic type-electromagnetic vertex  $\propto \chi_{ijk}E^P_{i}E^S_jE^I_k$, with $P,S,I$ for pump, signal and idler fields\cite{pdc1,pdc2,pdc3}. For an intense monochromatic laser beam, the pump field is usually taken to be classical whereas the signal and idler fields are quantized. The two photons emerging from the non-linear crystal (signal and idler) feature momentum and polarization entanglement as a consequence of the non-linearity similarly to the two photons emerging from axion decay. The down conversion is of type I or type II respectively if the photon pair features parallel or perpendicular polarizations.

In a finite sized crystal and during a finite time interval,  the momenta and frequencies of the signal and idler photons are determined by the \emph{sinc}-type functions\cite{pdc1}
\be \Pi^3_{m=1}\,\Bigg[ \frac{\sin\big[\frac{1}{2}\big(\vec{k}_P-\vec{k}_S-\vec{k}_I \big)_m L_m \big]}{\frac{1}{2}\big(\vec{k}_P-\vec{k}_S-\vec{k}_I \big)_m}\Bigg] \times \frac{\sin\big[\frac{1}{2}\big(\omega_P-\omega_S-\omega_I \big) t \big]}{\frac{1}{2}\big(\omega_P-\omega_S-\omega_I \big) } \,, \label{sincs} \ee which for large size in each direction $L_m$ and for large time $t$ yield the (approximate) phase and frequency matching conditions
\be \vec{k}_P \simeq  \vec{k}_S+\vec{k}_I ~~;~~ \omega_P\simeq \omega_S+\omega_I \,.\label{match}\ee These conditions must be compared to the momentum conservation condition in axion decay $\vec{k}_1+\vec{k}_2 = \vec{k}$ in the results of the previous section and the resonant denominators in the amplitude (\ref{c2gama}) which yield the largest amplitude for the energy conserving conditions $E_{2\gamma} \simeq \varepsilon^{\pm}$. In the limit of vanishing width these resonant denominators can be replaced by approximate energy conserving delta functions, the ``blurring'' of the energy conserving condition is a consequence of the uncertainty associated with the lifetime of the decaying state. The time dependent \emph{sinc} function in (\ref{sincs}) is the usual function arising in time dependent perturbation theory that yields energy conservation in the long time limit a la Fermi's golden rule. Therefore the kinematic constraints from parametric down conversion (phase and frequency matching) are equivalent to those in the two-photon final state
from axion decay, this is of course expected from the kinematics of energy-momentum conservation. The main difference is in the polarization entanglement. Whereas for axion decay, the polarization function consequence of polarization entanglement is given by eqn. (\ref{Cpola}), in parametric down conversion the dependence in the polarization is given by a function of the form\cite{pdc1}
\be \mathcal{C}(\vec{k}_S,\lambda_S;\vec{k}_I,\lambda_I) \propto \widetilde{\chi}_{ijl}\,\vec{\epsilon}_j(\vec{k}_S) \,\vec{\epsilon}_l(\vec{k}_I) \,,\label{cpolapdc}\ee where $\widetilde{\chi}$ is a frequency Fourier transform of the non-linear susceptibility.

These differences notwithstanding, the main remarkable similarity between axion decay and parametric down conversion is that of momentum and polarization entanglement in the final two photon state. In references\cite{pdc1,pdc2,pdc3} second order (intensity) interferometry, with (HBT) second order coherences,  has been proposed to study the correlations between signal and idler photons, in either type of parametric down conversion. And in ref.\cite{bi} polarization resolved (HBT) second order coherence was used to test violations of Bell inequalities in a two-photon correlation experiment from the interference of signal and idler photons from parametric down conversion.

 Therefore the main similarity between the case of condensed matter and cosmological axions and parametric down conversion bolsters the argument anticipated in the previous section that (HBT) second order coherence, perhaps with momentum and polarization resolution may be suitable probes for detection  of both synthetic and cosmological axions. In the case of hybridization these intensity correlations will feature quantum beats. Furthermore, even in the case when the cosmological coupling $g_c$ is so small so as to make the quantum beats unobservable, both momentum and polarization correlations in the two-photon final state are a distinct telltale of axionic excitations in topological insulators and in Weyl semimetals  which are imprinted in the second order (HBT) correlations.

Therefore, axionic collective excitations in condensed matter systems  may be probed  with intensity interferometry via    momentum and polarization resolved second order (HBT) cross correlations.

 \section{Discussion.}\label{sec:discussion}

  \vspace{1mm}

  \textbf{ Beyond the Markov approximation:} In our analysis, we have taken the Markovian limit in the form of the long-time limit in equation (\ref{pp}). This is ubiquitous and familiar in the Weisskopf-Wigner theory of spontaneous emission in single or multilevel atoms\cite{zubairy,meystre,ww,garraway,plenio}. Keeping the full time dependence entails a   time dependent $2\times 2$ Hamiltonian, and the solution of the effective Schroedinger equation involves the time ordered exponential of this Hamiltonian, with obvious technical complications. The time dependence of the Hamiltonian may affect the early time, transient dynamics, but for weak coupling we expect the intermediate and long time dynamics to be reliably described by the Markovian limit. Of course this expectation must be analyzed  and confirmed with a rigorous calculation, which is beyond the main objectives of this article and is postponed to further study.

  \vspace{1mm}

  \textbf{Renormalization of the real part of the self energy (Lamb shift):} The real part of the self-energy matrix $\Delta_{ab}$, given by equation (\ref{RIsig}) is   divergent in the limit of large ultraviolet cutoff scale $\Lambda$. The diagonal matrix elements $\Delta_{ss},\Delta_{cc}$ correspond to Lamb-shifts and are absorbed into a  renormalization of the single particle energies for  synthetic and cosmological axions,   as in equation (\ref{ERs}).
  However, the off-diagonal matrix elements cannot be absorbed by any terms in the (bare) Lagrangian as there are no bare terms in the Lagrangian density that mix the synthetic and cosmological excitations.  In principle, to renormalize these terms  the bare Lagrangian density must be extended to include local off-diagonal (mass-type) terms that mix synthetic and cosmological axions as \emph{counterterms} of order $g_s g_c$, which are then  chosen to cancel the divergences in the off-diagonal terms $\Delta_{sc},\Delta_{cs}$. In the non-degenerate case studied in the previous sections (and the most likely scenario), the contributions from the off-diagonal self-energy matrix elements  are subleading and their renormalization is not needed to leading order in the couplings as derived in the previous sections. However, in the (unlikely) case in which the energy differences are of the order of or smaller than the self-energy corrections, the full renormalization program including counterterms in the bare Lagrangian must be addressed. This theoretical possibility merits further and deeper study, again a technical aspect that is well beyond the realm of our objectives here.

  \vspace{1mm}

  \textbf{Nearly degenerate case:} There is a wide range of masses for the cosmological axion that are phenomenologically viable\cite{marsh,sikivie1,graham,irastorza}, $10^{-21}\,\mathrm{eV} \lesssim m_c \lesssim \mathrm{GeV}$ with the low end favored for ultra light dark matter, whereas the mass range for synthetic axions, typically determined by the material properties is within the $\mathrm{meV}$\cite{nomura,rundong}. Therefore, unless there is a   coincidence, synthetic and cosmological axions  are unlikely to be nearly degenerate, namely with a difference in masses of the same order as the self-energy corrections which are perturbatively small for small coupling. The results (\ref{c2gamand},\ref{tilG2finnd}) were obtained under the assumption that synthetic and cosmological axions are \emph{not nearly degenerate} so that the product of off-diagonal self-energy matrix elements in (\ref{wiD})   and the off diagonal coefficients in (\ref{wiR}) can be safely neglected.

  If both axions are nearly degenerate these contributions can no longer be neglected and the full dispersion relations of the (quasi) normal modes must be worked out in detail. While this by itself is an interesting but complicated theoretical exercise, we do not consider this possibility to be realistic and focused on the most likely, non-degenerate scenario to highlight the main physical phenomena of hybridization and its consequences.

\vspace{1mm}

\textbf{Induced synthetic axion condensate:}  A corollary of hybridization is that even if the initial amplitude of one of the axion species vanishes, it is generated upon time evolution, as described by equations (\ref{indusyn},\ref{inducos}). In the non-degenerate case, the main focus of the discussion in the previous sections, the coefficients $\widetilde{\beta};\widetilde{\gamma}$ in equations (\ref{indusyn},\ref{inducos}) are perturbatively small and of order $g_s g_c$, but are enhanced to be of $\mathcal{O}(1)$ in the nearly degenerate case. A current cosmological paradigm posits that dark matter is in the form of a spatially homogeneous but time dependent cosmological axion condensate\cite{marsh,graham,sikivie1}. Hybridization entails that such condensate   induces a condensate of axionic excitations in topological insulators or Weyl semimetals. Accordingly, this should translate into   (HBT) correlations and coincident photodetection with a characteristic polarization pattern even if  axionic collective excitations are not generated by the experimental setup. This is certainly an intriguing possibility for indirect detection of axionic dark matter which excites the collective axionic modes in these condensed matter platforms via the hybridization mechanism.

\vspace{1mm}

  \section{Conclusions and further questions}\label{sec:conclusions}

  Cosmological axions ($\phi_c$), a compelling dark matter candidate, and synthetic axions ($\phi_s$), emergent collective excitations in topological insulators and in Weyl semimetals couple to electromagnetism via a (pseudoscalar) Chern-Simons term. Their mutual coupling to electromagnetism implies that both species of axions  hybridize via the emission and absorption of two photons $\phi_c \Leftrightarrow 2\gamma \Leftrightarrow \phi_s$, resulting in an off-diagonal self-energy matrix.

  An analogy with a three-level ``V-type'' atomic system, identifying the two upper levels with the synthetic and cosmological axions decaying to a ground level via   two-photon emission  leads us to generalize and extend the Weisskopf-Wigner theory of spontaneous emission in multilevel atoms to study the dynamics of hybridization and its consequences. This formulation allows us to obtain the final two-photon state which features momentum and polarization entanglement and quantum beats as a consequence of the interference between the decay paths.

  A two photon correlated final state motivates us  to obtain a second order (HBT) coherence to study its properties via intensity interferometry with coincident two-photon photodetection. We introduce a momentum and polarization resolved second order (HBT) cross correlation which probes both momentum and polarization entanglement and features quantum beats from the interference between the decay paths. The distinct pattern of polarization entanglement is a consequence of the axion coupling to the pseudoscalar Chern-Simons term.

  An important corollary of hybridization is that a condensate of the cosmological axion, a main paradigm of theories of dark matter, induces a condensate of the synthetic axion, suggesting an indirect method of detecting the cosmological species from the excitation and decay of the condensed matter species, probed with (HBT) correlations.

  The limit of vanishing coupling of the cosmological axion allows us to extend the results to the case of synthetic axions independently, since the two-photon final state features the kinematic and polarization entanglement properties as consequences of the Chern-Simons coupling to electromagnetic fields individually for each species. Therefore we argue that momentum and polarization resolved (HBT) second order correlations are a useful probe to detect   axionic collective excitations  in condensed matter platforms.

  We point out a remarkable similarity to the phenomenon of parametric down conversion in non-linear crystals, with correlations between the down-converted idler and signal photons that can be probed via (HBT) interferometry, thereby suggesting the same type of experimental setups to study axionic excitations in condensed matter systems.

  In this theoretical study we did not address possible shortcomings and limitations arising from real materials and experimental realizations, nor have we discussed the various proposed materials as topological insulators or Weyl semimetals. We  focused solely on the theoretical underpinnings of   hybridization, its consequences such as quantum beats,  and how to implement a possible
   theoretical framework to study them using intensity interferometry and (HBT) correlations. Obviously, the feasibility of observation of the phenomena associated with hybridization and or the implementation of interferometric studies of momentum and polarization entanglement with coincident two-photon photodetection and (HBT) correlations must be assessed experimentally.

  A further avenue to study the correlations of the final two photon state is to obtain the expectation value  in this state of the quantum Stokes parameters introduced in references\cite{stokes1,stokes2,stokes3}.
  In these references typically a monochromatic beam is assumed to propagate along a particular direction, and quantum Stokes parameters are associated with the perpendicular polarizations. This framework must be adapted to the case of the two-photon final state emerging from axion decay, because such a state is not monochromatic and is a mixture of polarizations. These aspects and their possible implementations as probes of axions will be the focus of further study.

\acknowledgements
 The author thanks Shuyang Cao for discussions during the early stages of this work, and  gratefully acknowledges  support from the U.S. National Science Foundation through grants   NSF 2111743 and NSF 2412374.

\appendix

\section{Hamiltonian quantization with a Chern-Simons term}\label{app:hamcs}
Let us consider the   Lagrangian density of a  generic pseudoscalar axion field  $\phi(\vx,t)$ coupled solely to electromagnetism via a Chern-Simons term but in absence of external charge and current densities,
\be \mathcal{L} = \frac{1}{2}  \Big(\vec{E}^2(\vx,t)-\vec{B}^2(\vx,t)\Big) + g \phi(\vx,t)\,\vec{E}(\vx,t)\cdot \vec{B}(\vx,t) + \frac{1}{2}\Big[ \big(\frac{\partial}{\partial t}\phi(\vx,t)\big)^2  - v^2 \,\big( \vec{\nabla}\phi(\vx,t)\big)^2 - m^2 \phi^2(\vx,t)\,\Big]  \,,\label{lag}\ee where $v$ is a material dependent parameter in the case of synthetic axions and $v=1$ for the cosmological axion.
In terms of the vector and scalar potentials $\vec{A}(\vx,t);A_0(\vx,t)$ respectively,
\be \vec{E}(\vx,t) = -\frac{\partial}{\partial t}\vec{A}(\vx,t)-\vec{\nabla}  A_0(\vx,t)~~;~~ \vec{B}(\vx,t) = \vec{\nabla} \times \vec{A}(\vx,t) \,.\label{EBfields}\ee   The equations of axion electrodynamics follow from the variational principle, these are given by\cite{wilczekaxion}
\bea \vec{\nabla} \times \vec{E}(\vx,t) & = & - \frac{\partial}{\partial t} \vec{B}(\vx,t)\,, \label{curle}\\ \vec{\nabla}\cdot \vec{B}(\vx,t) & = & 0 \,,\label{divb}\\
\vec{\nabla}\cdot \vec{E}(\vx,t) & = & -g \vec{B}(\vx,t)\cdot \vec{\nabla} \phi(\vx,t) \,,\label{divE}\\
\vec{\nabla} \times \vec{B}(\vx,t) & = & \frac{\partial}{\partial t} \vec{E}(\vx,t) + g \,\Big( \vec{B}(\vx,t)\,\frac{\partial}{\partial t}\phi(\vx,t)+ \big(\vec{\nabla}\phi(\vx,t)\big) \times \vec{E}(\vx,t) \Big)\,,\label{curlb}
\eea and the equation of motion for the pseudoscalar field is
\be \frac{\partial^2}{\partial t^2} \phi(\vx,t)-v^2 \nabla^2\phi(\vx,t)+m^2 \phi(\vx,t)-g \vec{E}(\vx,t)\cdot\vec{B}(\vx,t) =0 \,.\label{eomfi}\ee

Including external charge and current densities modify the right hand side of  equations (\ref{divE},\ref{curlb})   by the usual source terms in electrodynamics. In passing to the Hamiltonian description one recognizes that the coupling via the Chern-Simons term modifies the canonical momentum conjugate to $\vec{A}(\vx,t)$, whereas, as usual the canonical momentum conjugate to $A_0(\vx,t)$ vanishes,
\be \vec{\Pi}(\vx,t) = \frac{\partial \mathcal{L}}{\partial \dot{\vec{A}}(\vx,t)} = -\vec{E}(\vx,t)- g\, \phi(\vx,t)\,\vec{B}(\vx,t)~~;~~ \Pi_0(\vx,t) =   \frac{\partial \mathcal{L}}{\partial \dot{A}_0(\vx,t)}=0 \,,\label{Pis}\ee and for the pseudoscalar field
\be \pi_{\phi}(\vx,t) = \frac{\partial}{\partial t} \phi(\vx,t)\,. \label{pifi}\ee
We note that equation (\ref{divE}) is not an equation of motion, as there are no time derivatives, but a constraint. In particular it yields the following constraint on the canonical momentum conjugate to the vector potential   (in absence of external sources)
\be \vec{\nabla}\cdot\vec{\Pi}(\vx,t) = 0\,.\label{Piconst}\ee
 The Hamiltonian density is obtained as usual
\be \mathcal{H} = \vec{\Pi}\cdot \frac{\partial}{\partial t}\vec{A}+ \pi_{\phi} \frac{\partial}{\partial t}\phi -\mathcal{L} \label{hamdens}\ee by expressing $\dot{\vec{A}},\dot{\phi}$ in terms of $\vec{\Pi},\pi_{\phi}$ respectively, yielding the total Hamiltonian
\be H = \int d^3 x \Bigg\{ \frac{1}{2} \vec{\Pi}^{\,2} + \frac{1}{2} \vec{B}^{\,2} - \vec{\Pi}\cdot \vec{\nabla}A_0 + g\,\phi \, \vec{\Pi}\cdot\vec{B}+ \frac{g^2}{2}\,\phi^2\,\vec{B}^2  + \frac{1}{2}\pi^2_{\phi} + \frac{v^2}{2} (\vec{\nabla}\phi)^2 + \frac{m^2}{2}\phi^2  \Bigg\}   \,.\label{tothamcs}\ee
Hamilton's equations of motion yield
\bea  \dot{\vec{A}}  &  =  &  \vec{\Pi} - \vec{\nabla}A_0 + g \phi \vec{B} \,,\label{adot}\\
\dot{\phi} &  = & \pi_{\phi}\,, \label{fidot}
\eea

which are precisely the definition of the canonical momenta  $\vec{\Pi},\pi_{\phi}$, equations (\ref{Pis},\ref{pifi}) respectively, and
\bea \dot{\vec{\Pi}}   &  =  &    -\big( \vec{\nabla}\times \vec{B}\big) - g^2 \,\Big\{2 \,\phi\,  \big( \vec{\nabla}\phi \times \vec{B} \big) + \phi^2 \, \big(\vec{\nabla}\times \vec{B}\big)  \Big\}-  g\,\Big\{\big( \vec{\nabla} \phi \times \vec{\Pi} \big) + \phi \big(\vec{\nabla}\times \vec{\Pi} \big) \Big\}\,, \label{Pidot1} \\
\dot{\pi}_{\phi} & = & v^2 \nabla^2\phi -m^2 \phi -g \vec{\Pi}\cdot \vec{B}-g^2 \phi \vec{B}^2  \,.  \label{dotpifi} \eea

Using the first relation in equation (\ref{Pis}),
 the equations of motion (\ref{curlb}, \ref{eomfi} ) are obtained  after straightforward algebra.  The equation (\ref{Piconst}) cannot be obtained from Hamilton's equations of motion as it is a constraint.
  Using canonical commutation relations, it follows that $\vec{\nabla}\cdot \vec{\Pi}$ is the generator of time independent U(1) gauge transformations, just as $\vec{\nabla}\cdot \vec{E}$ in quantum electrodynamics (QED) without sources, therefore gauge invariant states are annihilated by $\vec{\nabla}\cdot \Pi$. Therefore,   $\vec{\nabla}\cdot \vec{\Pi}=0$ is a constraint  that must be applied onto physical states, just as $\vec{\nabla}\cdot \vec{E} =0$ in quantum electrodynamics without sources. Upon integration by parts in (\ref{tothamcs}) and using the constraint (\ref{Piconst}), when acting on physical states, namely those annihilated by this constraint, the third term in the total Hamiltonian (\ref{tothamcs}) vanishes. It is convenient to work in Coulomb gauge
\be \vec{\nabla}\cdot \vec{A}(\vx,t) =0 \,,\label{coulomb}\ee wherein the constraint (\ref{Piconst}) becomes Poisson's equation with a source determined by the Chern-Simons term, namely
\be \nabla^2 A_0(\vx,t) = g \vec{B}(\vx,t)\cdot \vec{\nabla}\phi(\vx,t)\,.\label{poissoncs}\ee

Adapting the Hamiltonian framework to the case of synthetic and cosmic axions, it is simply a matter of replacing
\be g\phi \rightarrow (g_s \phi_s + g_c \phi_c)\,,\label{replacement}\ee in the above  results.

  \subsection{Interaction picture quantization:}\label{subsec:ipquan}

  Upon integration by parts of the third term in (\ref{tothamcs}) and in the physical (gauge invariant) sector, imposing the constraint (\ref{Piconst}), we write the Hamiltonian (\ref{tothamcs}) as

  \be H = H_0 +H_I\,\label{hspli}\ee where
  \bea H_0 & = & \int d^3 x \Bigg\{ \frac{1}{2} \vec{\Pi}^{\,2} + \frac{1}{2} \vec{B}^{\,2}    + \frac{1}{2}\pi^2_{\phi} + \frac{v^2}{2} (\vec{\nabla}\phi)^2 + \frac{m^2}{2}\phi^2  \Bigg\}\,\label{hzero} \\
  H_I & = &   \int d^3 x \Bigg\{    g\,\phi \, \vec{\Pi}\cdot\vec{B}+ \frac{g^2}{2}\,\phi^2\,\vec{B}^2    \Bigg\}\,.\label{hi} \eea Passing to the interaction picture of the free field Hamiltonian $H_0$,
  \be H_I(t) = e^{iH_0t} H_I e^{-iH_0t}\,. \label{hiip}\ee

  In interaction picture it follows that
  \be \vec{\Pi}(\vx,t) = \dot{\vec{A}}(\vx,t)~~;~~ \pi_{\phi}(\vx,t) = \dot{\phi}(\vx,t)\,,\label{ippis}\ee and the constraint (\ref{Piconst}) is automatically fulfilled in Coulomb gauge (\ref{coulomb}).   In the interaction picture the quantization of the (pseudo) scalar degrees of freedom, namely synthetic (s) and cosmological (c) axions is achieved via the
  expansion in the quantization volume $V$
  \be \phi_{s,c}(\vx,t) = \frac{1}{\sqrt{V}}\,\sum_{\vk} \frac{1}{\sqrt{2E_{s,c}(k)}}\,\Big[b_{s,c}(\vk)\,e^{-iE_{s,c}(k)t}\,e^{i\vk\cdot \vx} + b^\dagger_{s,c}(\vk)\,e^{iE_{s,c}(k)t}\,e^{-i\vk\cdot \vx}  \Big]\,,\label{scalarquant}\ee in Coulomb gauge, the vector potential is given by
  \be \vec{A}(\vx,t) =  \frac{1}{\sqrt{V}}\,\sum_{\vk}\sum_{\lambda=1,2}  \frac{\vec{\epsilon}_\lambda(\vk)}{\sqrt{2k}}\, \Big[a_{\lambda}(\vk)\,e^{-ik t}\,e^{i\vk\cdot \vx} + a^\dagger_{\lambda}(\vk)\,e^{ikt}\,e^{-i\vk\cdot \vx}  \Big]\,,\label{vecpot}\ee where the   polarization unit vectors $\vec{\epsilon}_\lambda(\vk)$ are chosen real and $\hat{\vk},\vec{\epsilon}_{1,2}(\vk)$ form a right handed triad with the properties
  \be \vec{\epsilon}_{1}(\vk)\times\vec{\epsilon}_{2}(\vk)= \hat{\vk}~~;~~\vec{\epsilon}_{2}(\vk)\times\hat{\vk}= \vec{\epsilon}_{1}(\vk)~~;~~\vec{\epsilon}_{1}(\vk)\times\hat{\vk}=-\vec{\epsilon}_{2}(\vk)~~;~~ \vec{\epsilon}_{1}(-\vk)= - \vec{\epsilon}_{1}(\vk)~~;~~\vec{\epsilon}_{2}(-\vk)= \vec{\epsilon}_{2}(\vk)\,. \label{polas}\ee The annihilation and creation operators obey the usual canonical commutation relations. The magnetic field is as usual
  \be \vec{B}(\vx,t) = \vec{\nabla} \times \vec{A}(\vx,t) \,,\label{Bfield}\ee  and  the canonical momentum  conjugate to the vector potential in Coulomb gauge and interaction picture is
  \be \vec{\Pi}(\vx,t) = \frac{\partial}{\partial t}\vec{A}(\vx,t)= -\vec{E}(\vx,t) \,,\label{Pican}\ee

  We recognize that the last term in (\ref{hi}) features ultraviolet divergences. The vacuum expectation value of the operator $\vec{B}^2$ features zero point divergences, which entail an ultraviolet divergent renormalization of the mass of the pseudoscalar field, and the vacuum expectation value of the operator $\phi^2$ also features zero point ultraviolet divergences which entail a divergent renormalization of the permeability of the vacuum. Hence, we define the normal ordered interaction in (\ref{hi}) as
  \be g^2 \phi^2 \vec{B}^2 \rightarrow g^2 :\phi^2: :\vec{B}^2:\,, \label{nor}\ee where the normal ordering prescription in the interaction picture, $:(\cdots) :$ corresponds to arranging the annihilation operators always to the right of creation operators. However, we will obtain the transition matrix elements to \emph{linear order} in $g_s$ and $g_c$, therefore we will only keep the linear order interaction Hamiltonian, which     in the interaction picture and to this order  is given by
  \be H_I(t) = -\int d^3 x \Big( g_s \phi_s(\vx,t)+g_c\phi_c(\vx,t)\Big)\vec{E}(\vx,t)\cdot\vec{B}(\vx,t)\,,\label{HI}\ee with all operators being in the interaction picture, wherein $\vec{E}$ is given by eqn. (\ref{Pican}).

\section{Calculation of the spectral density:}\label{app:spec}
Writing
\be \vk\cdot \vec{k}_1 = k k_1\cos(\theta)~~;~~ d^3 k_1 =  2\pi\, k^2_1 dk_1 d(\cos(\theta)\,\label{dotprod}\ee and defining
\be q\equiv |\vec{k}-\vec{k}_1| = \sqrt{k^2+k^2_1 - 2kk_1\cos(\theta)} \,\label{qdef}\ee the spectral density (\ref{rhoabfin}) becomes
\be \rho_{ab}(k_0;k) = \frac{g_a g_b}{32\pi^2 k \,\sqrt{E_a(k)E_b(k)}}\,\int^\infty_0 dk_1 \int^{q^+}_{q^-} \Big[(q+k_1)^2-k^2 \Big]^2\,\delta(k_0-k_1-q) \,dq\,,\label{rhosi}\ee
where
\be q^+ = (k_1+k) ~~;~~ q^- = |k-k_1| \,,\label{qpm}\ee using the delta function   constraint, yielding  $q+k_1 = k_0$ leads to
\be \rho_{ab}(k_0;k) = \frac{g_a g_b}{32\pi^2 k \,\sqrt{E_a(k)E_b(k)}}\,\Big(k^2_0-k^2\Big)^2\,\int^\infty_0 dk_1 \int^{q^+}_{q^-}  \delta(k_0-k_1-q) \,dq\,.\label{rhosita}\ee The q-integral is non-vanishing and  equal to one  if the conditions
\be  k_0 >0 ~~;~~ |k_1-k| \leq k_0-k_1 \leq k+k_1\,,\label{condis}\ee are fulfilled. A graphical analysis shows that these conditions are fulfilled, for
\be k_0 > k ~~;~~ k_- \leq k_1 \leq k_+  \ee  where
\be k_{\pm} = \frac{1}{2} (k_0 \pm k)\,, \label{kpm}\ee finally yielding the result (\ref{rhoabfinal}).

\section{First order cross correlation:}\label{app:1stcorr}
The calculation of the first order cross correlation (\ref{tilG1}) is lengthy but straightforward, the result is the following
\bea  \widetilde{G}^{(1)}(\vec{p}_1,s_1;\vec{p}_2,s_2;t) & = & \sum_{\lambda_2} C^*_{2\gamma}(\vec{p}_1,s_1;\vec{k}_2,\lambda_2;t)C_{2\gamma}(\vec{p}_1,s_2;\vec{k}_2,\lambda_2;t)\,\delta_{\vec{k}_2,\vec{k}-\vec{p}_1}\delta_{\vec{p}_1,\vec{p}_2}\nonumber \\
& + & \sum_{\lambda_1} C^*_{2\gamma}(\vec{p}_1,s_1;\vec{k}_1,\lambda_1;t)C_{2\gamma}(\vec{k}_1,\lambda_1;\vec{p}_1,s_2;t)\,\delta_{\vec{k}_1,\vec{k}-\vec{p}_1}\delta_{\vec{p}_1,\vec{p}_2}\nonumber\\
& + & \sum_{\lambda_2} C^*_{2\gamma}(\vec{k}_2,\lambda_2;\vec{p}_1,s_1;t)C_{2\gamma}(\vec{p}_1,s_2;\vec{k}_2,\lambda_2;t)\,\delta_{\vec{k}_2,\vec{k}-\vec{p}_1}\delta_{\vec{p}_1,\vec{p}_2}\nonumber\\
& + & \sum_{\lambda_1} C^*_{2\gamma}(\vec{k}_1,\lambda_1;\vec{p}_1,s_1;t)C_{2\gamma}(\vec{k}_1,\lambda_1;\vec{p}_1,s_2;t)\,\delta_{\vec{k}_1,\vec{k}-\vec{p}_1}\delta_{\vec{p}_1,\vec{p}_2}\label{tilG1fina}
\eea

\end{document}